\documentclass[aps,prd,preprint,groupedaddress,nofootinbib,eqsecnum]{revtex4}
\usepackage{amsmath} 
\usepackage{amsthm} 
\usepackage{amssymb}	
\usepackage{graphicx} 
\usepackage{slashed}
\usepackage{hyperref}
\usepackage{epstopdf}

\begin{document}

\author{David Chester}
\email{dchester@ucla.edu}
\affiliation{Mani Bhaumik Institute of Theoretical Physics,
Department of Physics and Astronomy,
UCLA, Los Angeles, CA 90095-1547, USA}
\title{The Radiative Double Copy for Einstein-Yang-Mills Theory}

\setcounter{page}{1}

\begin{abstract}
Recently, a double-copy formalism was used to calculate gravitational
radiation from classical Yang-Mills radiation solutions. This work shows that 
Yang-Mills theory coupled to a biadjoint scalar field admits a
radiative double copy that agrees with solutions in
Einstein-Yang-Mills theory at the lowest finite order. Within this context,
the trace-reversed metric $\bar{h}^{\mu\nu}$ is a natural 
double copy of the gauge boson $A^{\mu a}$. This work provides
additional evidence that solutions in gauge and gravity theories are
related, even though their respective Lagrangians and nonlinear equations of
motion appear to be different.
\end{abstract}

\maketitle

\tableofcontents

\section{Introduction}

The Lagrangians and equations of motion for gauge and gravity theories
appear to be rather different. Nevertheless, there are intriguing
double-copy connections between their solutions.  This includes the
Kawai-Lewellen-Tye (KLT) tree-level relations
between gauge and gravity amplitudes in string theory~\cite{Kawai:1985xq} and the Bern-Carrasco-Johansson
(BCJ) double-copy relations between diagrams in quantum field theory~\cite{Bern:2008qj}.  The
BCJ double-copy relations are based on color-kinematics duality, which
gives particularly simple constructions of gravity amplitudes
starting from gauge-theory amplitudes.

At tree level the BCJ amplitude relations are
proven~\cite{Bern:2010yg,Bern:2010ue,ArkaniHamed:2009dn,Chen:2011jxa,delaCruz:2015dpa}. Numerous calculations at higher
loops provide evidence for the loop-level double-copy
conjecture~\cite{Carrasco:2011hw,Carrasco:2011mn,Bern:2012uf,Bern:2013yya}
and progress has been made to understand analogous monodromy relations, 
extending KLT relations to loop level~\cite{BjerrumBohr:2009rd,Stieberger:2009hq,BjerrumBohr:2010zs,Tye:2010dd,He:2016mzd,He:2017spx}.
Einstein-Yang-Mills scattering amplitudes~\cite{Adamo:2015gia,delaCruz:2016gnm,Stieberger:2016lng,Du:2016wkt}
can also be found via the double copy~\cite{Cachazo:2014nsa,Nandan:2016pya,Schlotterer:2016cxa} using
the CHY formalism~\cite{Cachazo:2013gna}.
Biadjoint scalar fields can be used to find solutions in Yang-Mills~\cite{White:2016jzc}, and solutions in a Yang-Mills-biadjoint-scalar
theory have been shown to give scattering amplitudes in
Einstein-Yang-Mills~\cite{Chiodaroli:2014xia,Chiodaroli:2015rdg,Chiodaroli:2016jqw}.

With the recent experimental detection of gravitational waves by
LIGO~\cite{PhysRevLett.116.061102}, precision calculational tools for
gravitational wave emission are essential. Exploiting color-kinematics duality to relate
radiation solutions between Yang-Mills and general relativity is
attractive because general relativity is difficult to solve and the
double copy has been shown to work for a wide variety of gravity
theories~\cite{Anastasiou:2013hba,Anastasiou:2017nsz,Johansson:2017srf} The connection between
radiation solutions of gauge theory and gravity has been described
recently~\cite{Monteiro:2014cda,Ridgway:2015fdl,Luna:2015paa,Luna:2016due,Bahjat-Abbas:2017htu,Adamo:2017nia,Luna:2017dtq}.
The first example of using the radiative double copy to find nonlinear
terms in general relativity utilized perturbative Yang-Mills
solutions~\cite{Goldberger:2016iau}. Similarly a biadjoint scalar field 
can be used to find Yang-Mills radiation~\cite{Goldberger:2017frp}.

This work builds off the radiative double copy for
general relativity found by Goldberger et al.~\cite{Goldberger:2016iau} to find gravitational radiation in
Einstein-Yang-Mills theory. By comparing the differential equations of
the sources and fields in gauge theory and gravity, radiative diagrams are used to represent specific algebraic terms. 
Solutions in gravity can be found from Yang-Mills theory, and the diagrams with three-point vertices can be computed by stitching lower-order solutions together.
At leading order, the trace-reversed metric~\cite{MTW}, $\bar{h}^{\mu\nu}$, is a natural
double copy of the Yang-Mills potential $A^{\mu a}$~\cite{Chu:2016ngc}.
Motivation for a perturbative double copy can be seen at the
Lagrangian level, as the linearized gravity Lagrangian is quite
similar to the QED Lagrangian, a linearized version of the
Yang-Mills Lagrangian. Similarly, these two theories both have an analogous linearized wave equation. Remarkably, radiation solutions of nonlinear gauge and
gravity theories are related, at least when iterated perturbatively. A double copy of Yang-Mills-adjoint-scalar theory is also briefly mentioned, which can recover radiation solutions in Einstein-Maxwell theory.

While this paper focuses on classical solutions that could 
be calculated with more traditional methods~\cite{Will:1996zj,Flanagan:2005yc,Blanchet,Porto:2017dgs,Bernard:2017ktp,Bohe:2016gbl,Moxon:2017ozd,Damour:2016gwp,Bini:2017xzy,Damour:2017zjx}, the hope is that
the radiative double copy could help with difficult
calculations that may be more cumbersome to do in general
relativity alone. As more experimental data for gravitational
radiation is collected, new methods for calculating complicated
radiation processes are encouraged.

Section~\eqref{sec3} calculates radiation in Yang-Mills-biadjoint-scalar theory. Section~\eqref{sec4} calculates radiation in Einstein-Yang-Mills theory and the double copy is confirmed by direct calculation. Section~\eqref{sec5} states our concluding remarks. Appendix~\eqref{appendixB} calculates details of the gravitational contribution to the energy momentum pseudotensor and  Appendix~\eqref{AppRadRules} gives radiative Feynman rules for simple diagrams with three-point vertices.

\section{Radiation in Yang-Mills-Biadjoint-Scalar Theory}
\label{sec3}

\subsection{Equations of Motion and Initial Conditions}

In this section, the non-Abelian radiation field for Yang-Mills-biadjoint-scalar field theory is computed to first order in the weak-field approximation. 
To start, the Lagrangian associated with the Yang-Mills-biadjoint-scalar theory is
\begin{equation}
\mathcal{L} = -\frac{1}{4} F_{\mu\nu}^a F^{\mu\nu a} + \frac{1}{2}D_\mu \Phi^{\tilde{a}a}D^\mu \Phi^{\tilde{a}a} - \frac{y}{3}f^{abc}f^{\tilde{a}\tilde{b}\tilde{c}}\Phi^{\tilde{a}a}\Phi^{\tilde{b}b}\Phi^{\tilde{c}c},
\end{equation}
where $f^{abc}$ and $f^{\tilde{a}\tilde{b}\tilde{c}}$ refer to structure constants of different groups, the biadjoint scalar $\Phi^{\tilde{a}a}$ has an index associated with each gauge group, and $y=-ig\tilde{g}/2$ relates the conventions of Ref.~\cite{Chiodaroli:2014xia} with the conventions of Refs.~\cite{White:2016jzc,Goldberger:2017frp}. In principle, there could be an $\mathcal{O}(\Phi^4)$ term in the Lagrangian, but the coupling constant would have different dimensions than $y$ and is not needed for the double copy. The non-Abelian field strength is given by
\begin{equation}
F_{\mu\nu}^a(x) = \partial_\mu A_\nu^a(x) - \partial_\nu A_\mu^a (x) - gf^{abc}A_\mu^b(x) A_\nu^c(x),
\label{FieldStrength}
\end{equation}
and the the mostly minus metric will be used, such that $\eta^{\mu\nu} = \mbox{diag}(1,-1,-1,-1)$.
The covariant derivative is given by
\begin{equation}
D_\mu \Phi^{\tilde{a}a}(x) = \partial_\mu \Phi^{\tilde{a}a}(x) - g f^{abc} A_\mu^b(x) \Phi^{\tilde{a}c}(x).
\end{equation}

The equations of motion for the Yang-Mills field is 
\begin{equation}
D_\mu F^{\mu\nu a}(x) - g f^{abc} \Phi^{\tilde{a}b}(x) D^\nu\Phi^{\tilde{a}c}(x) = gJ^{\nu a}(x),
\end{equation}
where $J^{\mu a}(x)$ is a non-Abelian vector current acting as a source for the Yang-Mills field and is covariantly conserved, such that $D_\mu J^{\mu a} = 0$. The equation of motion for the biadjoint scalar field is
\begin{equation}
\partial_\mu D^\mu \Phi^{\tilde{a}a}(x) -gf^{abc}A_\mu^b(x) D^\mu\Phi^{\tilde{a}c}(x)  - yf^{abc}f^{\tilde{a}\tilde{b}\tilde{c}} \Phi^{\tilde{b}b}(x)\Phi^{\tilde{c}c}(x) = yJ^{\tilde{a}a}(x).
\end{equation}
For $N$ colliding charged particles, the worldline of particle $\alpha$ is $x_\alpha^\mu(\tau) = b_\alpha^\mu + v_\alpha^\mu \tau$ for $\tau\rightarrow -\infty$. These initial conditions specify an impact parameter $b_{\alpha\beta}^\mu = b_\alpha^\mu - b_\beta^\mu$ and a constant initial velocity $v_\alpha^\mu$ which satisfies $v_\alpha^2 = 1$. For arbitrary times near and after the collision,
\begin{equation}
x_\alpha^\mu(\tau) = b_\alpha^\mu + v_\alpha^\mu \tau + z_\alpha^\mu(\tau),
\end{equation}
where $z_\alpha^\mu(\tau)$ is the deflection due to the Yang-Mills and biadjoint scalar fields. The vector source for $N$ colliding charged particles is
\begin{equation}
\label{v-source}
J^{\mu a}(x) = \sum_{\alpha = 1}^N \int d\tau c_\alpha^a(\tau) v_\alpha^\mu(\tau)\delta^d(x-x_\alpha(\tau)),
\end{equation}
where $\alpha$ is a particle number label, $v_\alpha^\mu(\tau) = \frac{d x_\alpha^\mu(\tau)}{d\tau}$ is the velocity, and $c_\alpha^a(\tau)$ is the associated adjoint color charge~\cite{Sikivie}. The biadjoint source $J^{\tilde{a}a}(x)$ for $N$ particles is 
\begin{equation}
\label{s-source}
J^{\tilde{a}a}(x) = \sum_{\alpha = 1}^N \int d\tau c_\alpha^{\tilde{a}}(\tau) c_\alpha^a(\tau) \delta^d(x-x_\alpha(\tau)),
\end{equation}
where it is assumed that the color charges $c_\alpha^{\tilde{a}}(\tau)$ and $c_\alpha^a(\tau)$ are in two different gauge groups. 

The Lorenz gauge is taken by setting $\partial_\mu A^{\mu a} = 0$. In order to simplify these equations, the explicit dependence on the covariant derivatives is removed and gauge dependent sources $\hat{J}^{\mu a}$ and $\tilde{J}^{\tilde{a}a}$ are defined such that
\begin{equation}
\square A^{\mu a}(x) = g\hat{J}^{\mu a}(x), \mbox{ }\mbox{ }\mbox{ }\mbox{ }\square \Phi^{\tilde{a}a} = y\hat{J}^{\tilde{a}a},
\label{linEOM}
\end{equation}
where $\square \equiv \partial_\nu \partial^\nu$. With these definitions, the pseudovector source is
\begin{equation}
\hat{J}^{\mu a} = J^{\mu a} + f^{abc}\left[A_\nu^b(\partial^\nu A^{\mu c} + F^{\nu\mu c}) + \Phi^{\tilde{a}b}D^\mu\Phi^{\tilde{a}c}\right],
\end{equation}
where the pseudovector is locally conserved, $\partial_\mu \hat{J}^{\mu a} = 0$. The pseudoscalar source is given by
\begin{equation}
\hat{J}^{\tilde{a}a} = J^{\tilde{a}a} + \frac{g}{y}f^{abc}\left[ \partial_\mu(A^{\mu b}\Phi^{\tilde{a}c}) + A^{\mu b}D_\mu \Phi^{\tilde{a}c} + \frac{y}{g}f^{\tilde{a}\tilde{b}\tilde{c}}\Phi^{\tilde{b}b}\Phi^{\tilde{c}c}\right].
\end{equation}

Similar to the worldline $x_\alpha^\mu(\tau)$, the color charges are dynamical and are given initial conditions $c_\alpha^a(\tau) = c_\alpha^a$ and $c_\alpha^{\tilde{a}}(\tau) = c_\alpha^{\tilde{a}}$ for $\tau\rightarrow -\infty$. For times near and after the collision,
\begin{eqnarray}
c_\alpha^a(\tau) &=& c_\alpha^a + \bar{c}_\alpha^a(\tau), \nonumber \\
c_\alpha^{\tilde{a}}(\tau) &=& c_\alpha^{\tilde{a}} + \bar{c}_\alpha^{\tilde{a}}(\tau),
\end{eqnarray}
where $\bar{c}_\alpha^a(\tau)$ and $\bar{c}_\alpha^{\tilde{a}}(\tau)$ are the corrections due to the Yang-Mills and biadjoint scalar fields. The time evolution of the momentum is 
\begin{equation}
\frac{dp_\alpha^\mu(\tau)}{d\tau} =  gc_\alpha^a(\tau) F^{\mu\nu a}(x_\alpha(\tau)) v_{\alpha\nu}(\tau) -y\partial^\mu \Phi^{\tilde{a}a}(x_\alpha(\tau)) c_\alpha^a(\tau) c_\alpha^{\tilde{a}}(\tau),
\end{equation}
and the time evolution of the charges is
\begin{eqnarray}
\frac{dc_\alpha^a(\tau)}{d\tau} &=& gf^{abc}v_\alpha^\mu(\tau) A_\mu^b(x_\alpha(\tau)) c_\alpha^c(\tau) - yf^{abc}\Phi^{\tilde{b}b}(x_\alpha(\tau))c_\alpha^{\tilde{b}}(\tau)c_\alpha^c(\tau), \nonumber \\
\frac{dc_\alpha^{\tilde{a}}(\tau)}{d\tau}  &=&  - yf^{\tilde{a}\tilde{b}\tilde{c}}\Phi^{\tilde{b}b}(x_\alpha(\tau))c_\alpha^b(\tau)c_\alpha^{\tilde{c}}(\tau).
\end{eqnarray}
These summarize all of the equations needed to iteratively solve for radiation in Yang-Mills-biadjoint-scalar theory within a weak-field approximation.

\subsection{Solutions of the Radiation Fields}
For weak fields, the lowest-order sources can be found from the initial conditions. The pseudocurrents in momentum space are
\begin{eqnarray}
\hat{J}^{\mu a}(k)|_{\mathcal{O}(g^0)} &=& \sum_{\alpha=1}^N e^{ik\cdot b_\alpha}(2\pi)\delta(k\cdot v_\alpha) v_\alpha^a c_\alpha^a, \nonumber \\
\hat{J}^{\tilde{a} a}(k)|_{\mathcal{O}(y^0)} &=& \sum_{\alpha=1}^N e^{ik\cdot b_\alpha}(2\pi)\delta(k\cdot v_\alpha)  c_\alpha^{\tilde{a}}c_\alpha^a,
\end{eqnarray}
which can be utilized to find the Yang-Mills and biadjoint scalar fields to lowest order from Eq.~\eqref{linEOM}, giving
\begin{eqnarray}
A^{\mu a}(x)|_{\mathcal{O}(g^1)} &=& -g \sum_{\alpha = 1}^N\int_l (2\pi)\delta(l\cdot v_\alpha)\frac{e^{-il\cdot(x-b_\alpha)}}{l^2}v_\alpha^\mu c_\alpha^a, \nonumber \\
\Phi^{\tilde{a} a}(x)|_{\mathcal{O}(y^1)} &=& -y \sum_{\alpha = 1}^N\int_l (2\pi)\delta(l\cdot v_\alpha)\frac{e^{-il\cdot(x-b_\alpha)}}{l^2}c_\alpha^{\tilde{a}} c_\alpha^a.
\label{BiadjointFieldSolution}
\end{eqnarray}

The lowest-order fields can be used to find the deflections of the sources, given by
\begin{eqnarray}
\left. m_\alpha \frac{d^2 z_\alpha^\mu(\tau)}{d\tau^2}\right|_{\mathcal{O}(g^2)} &=& gc_\alpha^a\left(\partial^\mu A^{\nu a}(x_\alpha(\tau))|_{\mathcal{O}(g^1)} - \partial^\nu A^{\mu a}(x_\alpha(\tau))|_{\mathcal{O}(g^1)}\right)v_{\alpha \nu} ,\nonumber \\
\left. m_\alpha \frac{d^2 z_\alpha^\mu(\tau)}{d\tau^2}\right|_{\mathcal{O}(y^2)} &=& -y\partial^\mu\Phi^{\tilde{a}a}(x_\alpha(\tau))|_{\mathcal{O}(y^1)}c_\alpha^{\tilde{a}}c_\alpha^a.
\end{eqnarray}
Plugging in the derivatives of the lowest-order fields gives
\begin{eqnarray}
\label{z-corr}
\left. m_\alpha \frac{d^2 z_\alpha^\mu(\tau)}{d\tau^2}\right|_{\mathcal{O}(g^2)} &=& ig^2 \sum_{\beta\neq\alpha}(c_\alpha^a c_\beta^a)\int_l (2\pi)\delta(l\cdot v_\beta)\frac{e^{-il\cdot(b_{\alpha\beta}+v_\alpha\tau})}{l^2}\left[(v_\alpha\cdot v_\beta)l^\mu - (v_\alpha\cdot l)v_\beta^\mu\right], \nonumber \\
\left. m_\alpha \frac{d^2 z_\alpha^\mu(\tau)}{d\tau^2}\right|_{\mathcal{O}(y^2)} &=& -iy^2\sum_{\beta\neq\alpha}(c_\alpha^a c_\beta^a)c_\alpha^{\tilde{a}}c_\beta^{\tilde{a}}\int_l (2\pi)\delta(l\cdot v_\beta)\frac{e^{-il\cdot(b_{\alpha\beta}+v_\alpha\tau})}{l^2}l^\mu.
\end{eqnarray}
Note that writing the color charge contraction as $c_\alpha\cdot c_\beta$ would be ambiguous with our notation, as $c_\alpha^a c_\beta^a$ and $c_\alpha^{\tilde{a}}c_\beta^{\tilde{a}}$ are distinct. The first correction of the color charges to second order in $g$ is given by
\begin{eqnarray}
\left. \frac{d \bar{c}_\alpha^a(\tau)}{d\tau}\right|_{\mathcal{O}(g^2)} &=& gf^{abc}v_\alpha^\mu A_\mu^b(x_\alpha(\tau))|_{\mathcal{O}(g^1)}c_\alpha^c, \nonumber \\
\left. \frac{d \bar{c}_\alpha^a(\tau)}{d\tau}\right|_{\mathcal{O}(y^2)} &=& -yf^{abc}\Phi^{\tilde{b}b}(x_\alpha(\tau))|_{\mathcal{O}(y^1)}c_\alpha^{\tilde{b}}c_\alpha^c, \nonumber \\
\left. \frac{d \bar{c}_\alpha^{\tilde{a}}(\tau)}{d\tau}\right|_{\mathcal{O}(y^2)} &=& -yf^{\tilde{a}\tilde{b}\tilde{c}}\Phi^{\tilde{b}b}(x_\alpha(\tau))|_{\mathcal{O}(y^1)} c_\alpha^bc_\alpha^{\tilde{c}}.
\end{eqnarray}
Once again, plugging in the lowest-order fields gives
\begin{eqnarray}
\label{c-corr}
\left. \frac{d \bar{c}_\alpha^a(\tau)}{d\tau}\right|_{\mathcal{O}(g^2)} &=& -g^2\sum_{\beta\neq\alpha}f^{abc}c_\beta^bc_\alpha^c (v_\alpha\cdot v_\beta)\int_l(2\pi)\delta(l\cdot v_\beta)\frac{e^{-il\cdot(b_{\alpha\beta} + v_\alpha \tau)}}{l^2}, \nonumber \\
\left. \frac{d \bar{c}_\alpha^a(\tau)}{d\tau}\right|_{\mathcal{O}(y^2)} &=& y^2\sum_{\beta\neq\alpha}f^{abc} c_\beta^b c_\alpha^c c_\alpha^{\tilde{b}}c_\beta^{\tilde{b}}\int_l(2\pi)\delta(l\cdot v_\beta)\frac{e^{-il\cdot(b_{\alpha\beta} + v_\alpha \tau)}}{l^2}, \nonumber \\
\left. \frac{d \bar{c}_\alpha^{\tilde{a}}(\tau)}{d\tau}\right|_{\mathcal{O}(y^2)} &=& y^2\sum_{\beta\neq\alpha} f^{\tilde{a}\tilde{b}\tilde{c}}c_\beta^{\tilde{b}}c_\alpha^{\tilde{c}}c_\alpha^b c_\beta^b \int_l(2\pi)\delta(l\cdot v_\beta)\frac{e^{-il\cdot(b_{\alpha\beta} + v_\alpha \tau)}}{l^2}.
\end{eqnarray}
These deflections can be utilized to find the sources to next order, which give the lowest-order radiation fields. 

Taking the Fourier transform of Eq.~\eqref{v-source} and integrating over the delta function gives
\begin{equation}
J^{\mu a}(k) = \sum_{\alpha = 1}^N \int d\tau e^{ik\cdot x_\alpha(\tau)} \left(v_\alpha^\mu + \frac{dz_\alpha^\mu(\tau)}{d\tau}\right)\left( c_\alpha^a + \bar{c}_\alpha^a(\tau) \right),
\end{equation}
Expanding these results perturbatively in $g$ and $y$ gives
\begin{equation}
J^{\mu a}(k) = \sum_{\alpha = 1}^N \int d\tau e^{ik\cdot (b_\alpha + v_\alpha\tau)}\left((1+ik\cdot z_\alpha)v_\alpha^\mu c_\alpha^a + v_\alpha^\mu\bar{c}_\alpha^a + \frac{dz_\alpha^\mu}{d\tau}c_\alpha^a\right) + \mathcal{O}(g^2y^2), 
\end{equation}
where explicit $\tau$ dependence has been suppressed and only terms to second order in $g$ or $y$ are kept. Integrating Eqs.~\eqref{z-corr} and \eqref{c-corr} allows for the second-order current to be found, which has Yang-Mills and biadjoint scalar contributions given by
\begin{eqnarray}
J^{\mu a}(k)|_{\mathcal{O}(g^2)} &=&  g^2\sum_{\substack{\alpha=1\\ \beta \neq\alpha}}^N \int_{l_\alpha,l_\beta} \mu_{\alpha,\beta}(k)\frac{l_\alpha^2}{k\cdot v_\alpha}\left[ if^{abc}c_\alpha^b c_\beta^c(v_\alpha\cdot v_\beta)v_\alpha^\mu \right.\\
&& \left. \null + \frac{c_\alpha^b c_\beta^b}{m_\alpha}c_\alpha^a\left\{ -v_\alpha\cdot v_\beta\left(l_\beta^\mu - \frac{k\cdot l_\beta}{k\cdot v_\alpha}v_\alpha^\mu\right) + k\cdot v_\alpha v_\beta^\mu-k\cdot v_\beta v_\alpha^\mu\right\}\right], \nonumber\\
J^{\mu a}(k)|_{\mathcal{O}(y^2)} &=& y^2\sum_{\substack{\alpha=1\\ \beta\neq\alpha}}^N c_\alpha^{\tilde{a}}c_\beta^{\tilde{a}}\int_{l_\alpha,l_\beta}\mu_{\alpha,\beta}(k)\frac{l_\alpha^2}{k\cdot v_\alpha}\left[\frac{c_\alpha^b c_\beta^b}{m_\alpha}c_\alpha^a\left(l_\beta^\mu - \frac{k\cdot l_\beta}{k\cdot v_\alpha}v_\alpha^\mu\right) - if^{abc}c_\alpha^b c_\beta^c v_\alpha^\mu \right], \nonumber
\end{eqnarray}
where an extra integral over $l_\alpha$ was added with a momentum conserving delta function such that $k = l_\alpha + l_\beta$ and
\begin{equation}
\mu_{\alpha,\beta}(k) = \left[ (2\pi)\delta(v_\alpha\cdot l_\alpha)\frac{e^{il_\alpha\cdot b_\alpha}}{l_\alpha^2}\right]\left[ (2\pi)\delta(v_\beta\cdot l_\beta)\frac{e^{il_\beta\cdot b_\beta}}{l_\beta^2}\right](2\pi)^d \delta^{d}(k-l_\alpha - l_\beta).
\end{equation}
The nonlinear field contributions to the pseudovector are represented by $j^{\mu a}$, which gives the following second order contributions
\begin{eqnarray}
j^{\mu a}(k)|_{\mathcal{O}(g^2)} &=& g^2\sum_{\substack{\alpha=1\\ \beta\neq\alpha}}^N if^{abc}c_\alpha^b c_\beta^c\int_{l_\alpha,l_\beta}\mu_{\alpha,\beta}(k)\left[2k\cdot v_\beta v_\alpha^\mu - v_\alpha\cdot v_\beta l_\alpha^\mu \right], \nonumber \\
j^{\mu a}(k)|_{\mathcal{O}(y^2)} &=& y^2\sum_{\substack{\alpha=1\\ \beta\neq\alpha}}^N if^{abc}c_\alpha^b c_\beta^c c_\alpha^{\tilde{a}}c_\beta^{\tilde{a}}\int_{l_\alpha,l_\beta}\mu_{\alpha,\beta}(k)l_\alpha^\mu.
\label{diag1f}
\end{eqnarray}

\begin{figure}
\includegraphics[scale=.8]{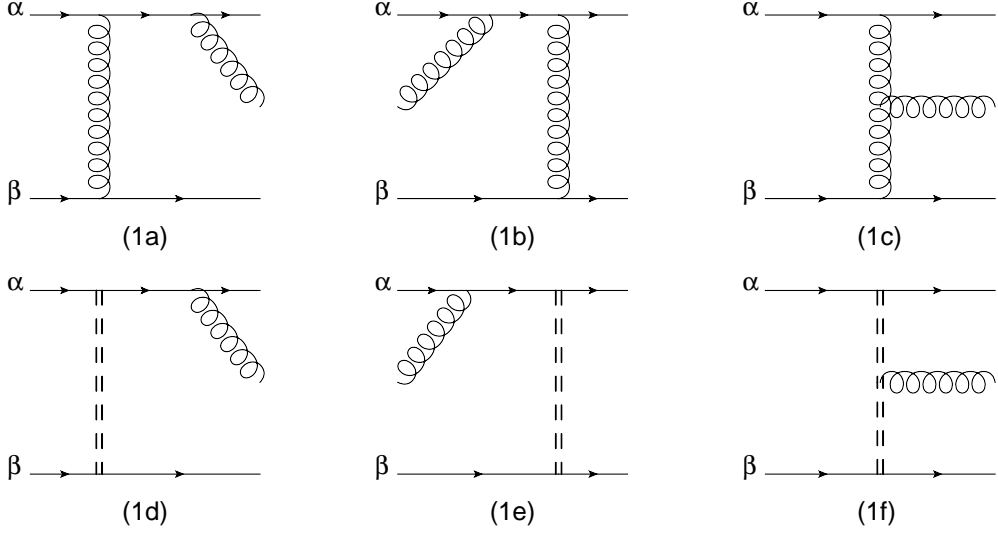}
\caption{These diagrams represent all of the contributions to $\hat{J}^{\mu a}$ in Yang-Mills-biadjoint-scalar theory. Straight lines represent matter fields, curly lines represent Yang-Mills fields, and doubly-dashed lines represent biadjoint scalar fields.}
\label{figYMbiadj}
\end{figure}

While $J^{\mu a}$ and $j^{\mu a}$ were computed algebraically, they also can be represented diagrammatically. Fig.~\eqref{figYMbiadj} depicts the diagrams associated with $\hat{J}^{\mu a}$ to second order in $g$ and $y$. The six diagrams are defined as
\begin{eqnarray}
(1\mbox{a})^{\mu a}(k) &\equiv& \sum_{\alpha = 1}^N \int d\tau e^{ik\cdot (b_\alpha + v_\alpha\tau)}\left(v_\alpha^\mu\bar{c}_\alpha^a(\tau)|_{\mathcal{O}(g^2)} + c_\alpha^a\left.\frac{dz_\alpha^\mu(\tau)}{d\tau}\right|_{\mathcal{O}(g^2)}\right),\nonumber \\
(1\mbox{b})^{\mu a}(k) &\equiv& \sum_{\alpha = 1}^N \int d\tau e^{ik\cdot (b_\alpha + v_\alpha\tau)}ik\cdot z_\alpha(\tau)|_{\mathcal{O}(g^2)}v_\alpha^\mu c_\alpha^a,\nonumber \\
(1\mbox{c})^{\mu a}(k) &\equiv& j^{\mu a}(k)|_{\mathcal{O}(g^2)} ,\nonumber \\
(1\mbox{d})^{\mu a}(k) &\equiv& \sum_{\alpha = 1}^N \int d\tau e^{ik\cdot (b_\alpha + v_\alpha\tau)}\left(v_\alpha^\mu\bar{c}_\alpha^a(\tau)|_{\mathcal{O}(y^2)} + c_\alpha^a\left.\frac{dz_\alpha^\mu(\tau)}{d\tau}\right|_{\mathcal{O}(y^2)} \right),\nonumber \\
(1\mbox{e})^{\mu a}(k) &\equiv& \sum_{\alpha = 1}^N \int d\tau e^{ik\cdot (b_\alpha + v_\alpha\tau)}ik\cdot z_\alpha(\tau)|_{\mathcal{O}(y^2)}v_\alpha^\mu c_\alpha^a,\nonumber \\
(1\mbox{f})^{\mu a}(k) &\equiv& j^{\mu a}(k)|_{\mathcal{O}(y^2)},
\label{matterRad}
\end{eqnarray}
where diagrams (1a), (1b), (1d), and (1e) give $J^{\mu a}(k)$ and diagrams (1c) and (1f) give $j^{\mu a}(k)$, both to second order in $g$ and $y$. The source $J^{\mu a}(k)$ was split into two types of diagrams, as (1a) represents radiation that was emitted after the particle was deflected, while (1b) represents radiation that was emitted before the particle was deflected. As such, it is anticipated that (1b) and (1e) should be proportional to the undeflected quantities $v_\alpha^\mu c_\alpha^a$, while (1a) and (1d) are in terms of corrections such as $\frac{d z_\alpha^\mu}{d\tau}c_\alpha^a$ and $v_\alpha^\mu \bar{c}_\alpha^a$. Diagrams (1c) and (1f) are computed in Appendix~\eqref{AppRadRules} from the three-point vertex with three vectors and the three-point vertex with two scalars and one vector, respectively. The six diagrams sum to give $\hat{J}^{\mu a}$ and satisfy the Ward identity $k_\mu \hat{J}^{\mu a}(k) = 0$. 

Summing up the three diagrams (1a)--(1c) is algebraically equivalent to $\hat{J}^{\mu a}(k)|_{\mathcal{O}(g^2)}$, giving
\begin{eqnarray}
\hat{J}^{\mu a}(k)|_{\mathcal{O}(g^2)} &=& g^2\sum_{\substack{\alpha=1\\ \beta\neq\alpha}}^N\int_{l_\alpha,l_\beta}\mu_{\alpha,\beta}(k)\left[if^{abc}c_\alpha^b c_\beta^c\left\{2(k\cdot v_\beta)v_\alpha^\mu + (v_\alpha\cdot v_\beta)\left(\frac{l_\alpha^2}{k\cdot v_\alpha}v_\alpha^\mu-l_\alpha^\mu\right)\right\}\right. \nonumber \\
&&\left. \null +\frac{c_\alpha^b c_\beta^b}{m_\alpha}\frac{l_\alpha^2c_\alpha^a}{k\cdot v_\alpha}\left\{v_\alpha\cdot v_\beta\left(\frac{k\cdot l_\beta}{k\cdot v_\alpha}v_\alpha^\mu - l_\beta^\mu\right)+k\cdot v_\alpha v_\beta^\mu - k\cdot v_\beta v_\alpha^\mu\right\}\right].
\label{YMradSolution}
\end{eqnarray}
which is the pure Yang-Mills contribution found by Ref.~\cite{Goldberger:2016iau}. Summing up the three diagrams (1d)--(1f) is equivalent to $\hat{J}^{\mu a}(k)|_{\mathcal{O}(y^2)}$, giving
\begin{eqnarray}
\hat{J}^{\mu a}(k)|_{\mathcal{O}(y^2)} &=& -y^2\sum_{\substack{\alpha=1\\ \beta\neq\alpha}}^N  c_\alpha^{\tilde{a}}c_\beta^{\tilde{a}}\int_{l_\alpha,l_\beta}\mu_{\alpha,\beta}(k) \nonumber \\
&&\null \times \left[\frac{c_\alpha^b c_\beta^b}{m_\alpha}\frac{l_\alpha^2c_\alpha^a}{k\cdot v_\alpha}\left(\frac{k\cdot l_\beta}{k\cdot v_\alpha}v_\alpha^\mu - l_\beta^\mu\right) + if^{abc}c_\alpha^b c_\beta^c\left(\frac{l_\alpha^2}{k\cdot v_\alpha}v_\alpha^\mu - l_\alpha^\mu\right)\right].
\label{radBiadjoint}
\end{eqnarray}
The radiative field must be gauge invariant and the above expression satisfies the Ward identity $k_\mu \hat{J}^{\mu a}(k)|_{\mathcal{O}(gy^2)} = 0$, as the identity must be satisfied order by order. Adding the above contributions to Eq.~\eqref{YMradSolution} gives the total source, $\hat{J}^{\mu a}$. To find the radiation field $A_{\textrm{rad}}^{\mu a}$ from the source $\hat{J}^{\mu a}$~\cite{jackson,Goldberger:2016iau}, 
\begin{equation}
A_{\textrm{rad}}^{\mu a}(x) = \frac{g}{4\pi r}\int\frac{d\omega}{2\pi} e^{-i\omega t}\hat{J}^{\mu a}(k),
\label{YMradField}
\end{equation}
where $k^\mu = \omega(1,\vec{x}/r)$.

\section{Gravitational Radiation in Einstein-Yang-Mills Theory}
\label{sec4}

\subsection{Equations of Motion and Initial Conditions}
The action for the Einstein-Yang-Mills-dilaton theory in consideration is
\begin{equation}
\label{gravAction}
S = \int d^dx \sqrt{-g}\left[-\frac{2}{\kappa^2}R - \frac{1}{4}g^{\mu\rho}g^{\nu\sigma}F_{\mu\nu}^{\tilde{a}}F_{\rho\sigma}^{\tilde{a}}+ \frac{2}{\kappa^2}(d-2)g^{\mu\nu}\partial_\mu\phi\partial_\nu\phi\right] - m\int d\tau e^\phi,
\end{equation}
where $\phi$ is the dilaton field and $d\tau = \sqrt{g_{\mu\nu}dx^\mu dx^\nu}$. By varying the action above, the energy-momentum pseudotensor contributions from the Yang-Mills field and the dilaton are given by
\begin{eqnarray}
8\pi G T_{\mu\nu} &=& R_{\mu\nu} - \frac{1}{2}g_{\mu\nu} R -(d-2)\left(\partial_\mu\phi \partial_\nu\phi - \frac{1}{2}g_{\mu\nu}g^{\rho\sigma}\partial_\rho\phi\partial_\sigma\phi\right) \nonumber \\
&&  \null + 8\pi G\left(g^{\rho\sigma}F_{\mu\rho}^{\tilde{a}}F_{\nu\sigma}^{\tilde{a}} - \frac{1}{4}g_{\mu\nu}g^{\rho\sigma} g^{\lambda\tau}F_{\rho\lambda}^{\tilde{a}}F_{\sigma\tau}^{\tilde{a}}\right).
\label{EFE}
\end{eqnarray}
According to Dirac, $\sqrt{|g|}T^{\mu\nu}$ is the density and flux of energy and momentum for matter~\cite{Dirac} such that in the presence of gravity, $N$ partices contribute
\begin{equation}
\sqrt{|g|}T^{\mu\nu}(x) = \sum_{\alpha = 1}^N m_\alpha\int d\tau v_\alpha^\mu(\tau) v_\alpha^\nu(\tau) \delta^d(x-x_\alpha(\tau)).
\end{equation}
A weak-field approximation is taken by introducing $h^{\mu\nu}$ as
\begin{eqnarray}
g_{\mu\nu} &=& \eta_{\mu\nu} + \kappa h_{\mu\nu}, \nonumber\\
g^{\mu\nu} &=& \eta^{\mu\nu} -\kappa h^{\mu\nu} + \kappa^2 h^{\mu\rho}h_{\rho}^\nu + \dots, \nonumber \\
|g| &\equiv& -\mbox{det}(g_{\mu\nu}) = 1 + \kappa h -\frac{\kappa^2}{2}(h^{\mu\nu} h_{\mu\nu}-h^2) + \dots,
\end{eqnarray}
where $h\equiv h_\rho^\rho$ and the radiation can be calculated perturbatively in powers of $\kappa$. Textbook presentations of gravitational waves often focus on linearized gravity~\cite{MTW}, which introduces the trace-reversed metric,
\begin{equation}
\bar{h}_{\mu\nu} \equiv h_{\mu\nu} - \frac{1}{2}\eta_{\mu\nu} h,
\end{equation}
and find that $\Box \bar{h}^{\mu\nu} = -\frac{\kappa}{2}T^{\mu\nu}$. If an effective energy-momentum pseudotensor $\hat{T}^{\mu\nu}$ was found to contain contributions from matter, nonlinear gravitational field contributions, and the other fields, then the following equation of motion can be solved iteratively within the context of the weak-field approximation
\begin{equation}
\Box \bar{h}^{\mu\nu} = -\frac{\kappa}{2} \hat{T}^{\mu\nu}.
\label{pseudoEOM}
\end{equation}
Due to the harmonic gauge condition, the pseudotensor satisfies $\partial_\mu\hat{T}^{\mu\nu} = 0$. The field contributions to the pseudotensor $t^{\mu\nu} \equiv \hat{T}^{\mu\nu} - T^{\mu\nu}$ will be found by expanding Eq.~\eqref{EFE}. The pseudotensor slightly differs from the common pseudotensor used by Landau and Lifshitz~\cite{LLpseudo,MTW,Blanchet} and is closer to ones used previously by Einstein and Dirac, giving
\begin{equation}
\hat{T}^{\mu\nu} = T^{\mu\nu} + t^{\mu\nu} \equiv \sqrt{|g|}T^{\mu\nu} + \hat{t}^{\mu\nu},
\end{equation}
where $\hat{t}^{\mu\nu}$ is conveniently defined to absorb $(1-\sqrt{|g|})T^{\mu\nu}$. In this section, the algebraic method of perturbing Einstein's field equations and iteratively solving for the radiation field is presented, leaving some technical details of the calculation of $\hat{t}^{\mu\nu}$ to Appendix \eqref{appendixB}. Since the three-point graviton vertex is derived from the Lagrangian of the full theory, diagrams can encode how to find higher order field contributions from linearized field solutions. In Appendix \eqref{AppRadRules}, radiative Feynman rules are provided for the diagrams ontributing to $t^{\mu\nu}$.

The Christoffel symbol $\Gamma^\rho_{\textrm{ }\textrm{ }\mu\nu}$ and the Ricci tensor $R_{\mu\nu}$ are given by
\begin{eqnarray}
\Gamma^\rho_{\textrm{ }\textrm{ }\mu\nu} &=& \frac{1}{2} g^{\rho\sigma}(g_{\sigma\nu,\mu}+g_{\sigma\mu,\nu} - g_{\mu\nu,\sigma}), \nonumber \\
R_{\mu\nu} &=& \Gamma^\rho_{\textrm{ }\textrm{ }\mu\nu,\rho} - \Gamma^{\rho}_{\textrm{ }\textrm{ }\mu\rho,\nu} + \Gamma^{\rho}_{\textrm{ }\textrm{ }\sigma\rho}\Gamma^\sigma_{\textrm{ }\textrm{ }\mu\nu} - \Gamma^\rho_{\textrm{ }\textrm{ }\sigma\nu}\Gamma^\sigma_{\textrm{ }\textrm{ }\mu\rho}.
\end{eqnarray}
After expanding the metric perturbatively in $\kappa$ and applying the gauge condition $\partial^\mu h_{\mu\nu} = \frac{1}{2}\eta_{\mu\nu}h^{,\mu}$,
\begin{eqnarray}
\Gamma^{\rho}_{\textrm{ }\textrm{ }\mu\nu} &=& \frac{\kappa}{2}\left(h^\rho_{\nu,\mu} + h^\rho_{\mu,\nu} - h_{\mu\nu,}^{\textrm{ }\textrm{ }\textrm{ }\textrm{ }\rho} -\kappa h^{\rho\sigma}(h_{\sigma\nu,\mu}+h_{\sigma\mu,\nu} - h_{\mu\nu,\sigma}) \right) + \mathcal{O}(\kappa^3),  \\
R_{\mu\nu} &=& -\frac{\kappa}{2}\Box h_{\mu\nu} + \frac{\kappa^2}{2}\left[h^{\rho\sigma}(h_{\mu\nu,\rho\sigma}+h_{\rho\sigma,\mu\nu} - h_{\sigma\nu,\mu\rho} - h_{\mu\rho,\sigma\nu}) \right.\nonumber \\
&& \left.\null + h_{\mu\rho,\sigma}h_\nu^{\rho,\sigma} - h_{\mu\rho,\sigma}h_\nu^{\sigma,\rho} + \frac{1}{2}h_{\rho\sigma,\mu}h^{\rho\sigma,}_{\textrm{ }\textrm{ }\textrm{ }\textrm{ }\nu} \right] + \mathcal{O}(\kappa^3). 
\end{eqnarray}
This gives the Ricci scalar $R$,
\begin{equation}
R = (\eta^{\mu\nu} - \kappa h^{\mu\nu})R_{\mu\nu} = -\frac{\kappa}{2}\Box h +\kappa^2\left(h^{\rho\sigma}\Box h_{\rho\sigma} + \frac{3}{4} h^{\rho\sigma,\mu}h_{\rho\sigma,\mu} - \frac{1}{2} h^{\mu\rho,\sigma}h_{\mu\sigma,\rho}\right) + \mathcal{O}(\kappa^3),
\end{equation}
To lowest order, $R^{\mu\nu} - \frac{1}{2}g^{\mu\nu} R \approx -\frac{\kappa}{2}\bar{h}^{\mu\nu}$, and all higher order terms in Eq.~\eqref{EFE} are subtracted to the other side of the equation to be absorbed into the definition of $t^{\mu\nu}$. Splitting these terms between $t^{\mu\nu}|_{\Delta h}$, $t^{\mu\nu}|_{\Delta \phi}$, and $t^{\mu\nu}|_{\Delta A}$ gives
\begin{eqnarray}
t^{\mu\nu}|_{\Delta h} &=& 2h_{\rho\sigma}\left(h^{\mu\rho,\nu\sigma} + h^{\nu\sigma,\mu\rho} - h^{\mu\nu,\rho\sigma} - h^{\rho\sigma,\mu\nu} \right) + h^{\mu\nu}\Box h - 2h^{\mu\rho}\Box h_\rho^\nu - 2h^{\nu\rho}\Box h_\rho^\mu  \nonumber \\
&& \null -2h^{\mu\rho,\sigma}\left(h^\nu_{\rho,\sigma} - h^\nu_{\sigma,\rho}\right) - h^{\rho\sigma,\mu}h_{\rho\sigma,}^{\textrm{ }\textrm{ }\textrm{ }\textrm{ }\nu} + \eta^{\mu\nu}\left[2h^{\rho\sigma}\Box h_{\rho\sigma} + h_{\rho\sigma,\lambda}\left(\frac{3}{2}h^{\rho\sigma,\lambda} - h^{\rho\lambda,\sigma}\right)\right], \nonumber \\
t^{\mu\nu}|_{\Delta\phi} &=& (d-2)\left(\frac{2}{\kappa}\right)^2\left(\partial^\mu \phi \partial^\nu \phi - \frac{1}{2}\eta^{\mu\nu}\partial_\rho \phi \partial^\rho\phi\right), \nonumber \\
t^{\mu\nu}|_{\Delta A} &=& -F^{\mu\rho\tilde{a}}F^{\nu\textrm{ }\tilde{a}}_{\textrm{ }\rho}+\frac{1}{4}\eta^{\mu\nu}F^{\rho\sigma \tilde{a}}F^{\tilde{a}}_{\rho\sigma}, 
\label{EYMdiagram}
\end{eqnarray}
where it is important to raise the indices on $R_{\mu\nu}-\frac{1}{2}g_{\mu\nu}R$ with $g^{\mu\nu}$ to get all of the necessary terms.

Similar to the previous section, the position of the particle is given by $x_\alpha^\mu(\tau)$, which has deflections $z_\alpha^\mu(\tau)$ which must be calculated from the field solutions. The matter is also assumed to have a color charge $c_\alpha^{\tilde{a}}(\tau)$, but their corrections do not source the lowest-order gravitational radiation field. The Christoffel symbol can be used to find the force on each particle, giving
\begin{equation}
\left. m_\alpha \frac{d^2 z_\alpha^\mu(\tau)}{d\tau^2}\right|_{\Delta h} = -\Gamma^\mu_{\textrm{ }\textrm{ }\nu\rho}m_\alpha v_{\alpha}^\nu v_\alpha^\rho.
\label{GRdeflection}
\end{equation}
The equation of motion utilized for the dilaton is
\begin{equation}
\left. m_\alpha \frac{d^2 z_\alpha^\mu(\tau)}{d\tau^2}\right|_{\Delta\phi} = m_\alpha v_{\alpha \nu}\partial^\mu \phi v_\alpha^\nu.
\label{dilatonDeflection}
\end{equation}
While this equation differs slightly from Ref.~\cite{Goldberger:2016iau}, both of our total pseudotensors agree and are the physical object that satisfies the gauge-invariant Ward identity.  The force due to the gauge field is
\begin{equation}
\left. m_\alpha\frac{d^2z_\alpha^\mu(\tau)}{d\tau^2}\right|_{\Delta A} = \tilde{g}c_\alpha^{\tilde{a}}F^{\mu\nu \tilde{a}}v_{\alpha\nu}(\tau).
\label{YMdeflection}
\end{equation}

\subsection{Solutions of the Radiation Fields}

\begin{figure}
\includegraphics[scale=.8]{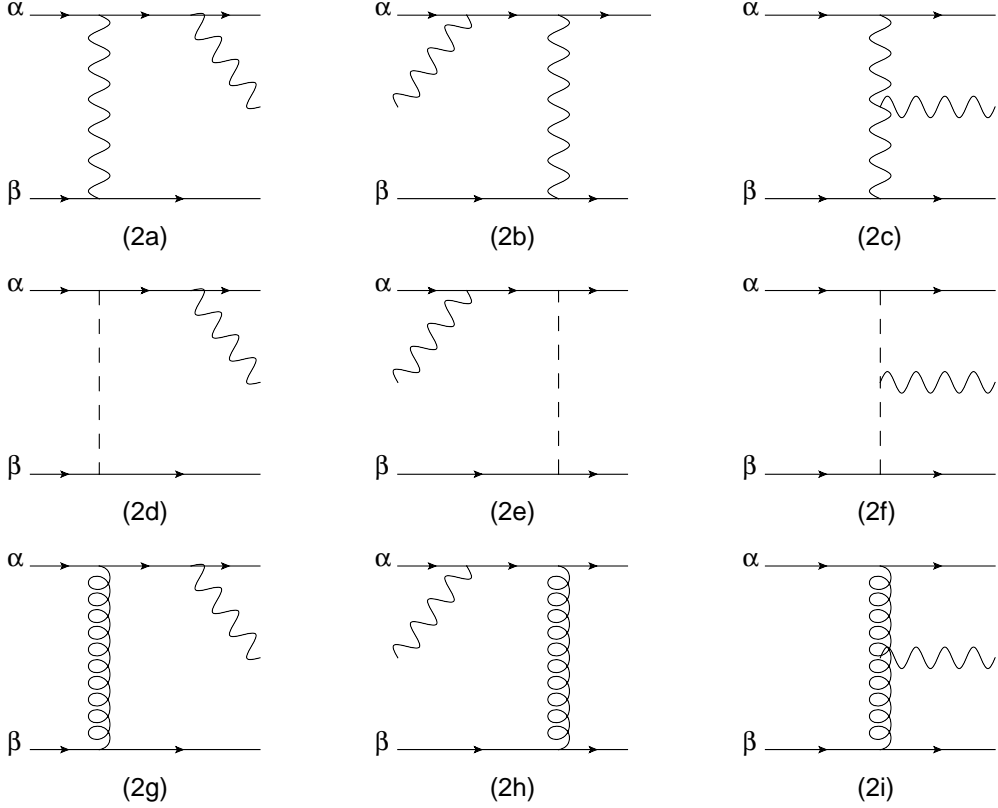}
\caption{These diagrams represent all of the contributions to $\hat{T}^{\mu\nu}$ in Einstein-Yang-Mills theory. The wavy lines represent gravitational fields, the dashed lines represent dilaton fields and the curly lines represent Yang-Mills fields.}
\label{figEYM}
\end{figure}

Fig.~\eqref{figEYM} shows nine diagrams which contribute to gravitational radiation for Einstein-Yang-Mills theory. Algebraically, the first three diagrams for the pure gravity contributions are
\begin{eqnarray}
(\mbox{2a})^{\mu\nu} + (2b)^{\mu\nu} &=& \sqrt{|g|}T^{\mu\nu}|_{\Delta h,\mathcal{O}(\kappa^2)}, \nonumber \\
(\mbox{2c})^{\mu\nu} &=& \hat{t}^{\mu\nu}|_{\Delta h,\mathcal{O}(\kappa^2)} \equiv t^{\mu\nu}|_{\Delta h,\mathcal{O}(\kappa^2)} + \left(1-\sqrt{|g|}\right)T^{\mu\nu}|_{\Delta h,\mathcal{O}(\kappa^2)},
\end{eqnarray}
while the diagrams with internal dilatons algebraically represent
\begin{eqnarray}
(\mbox{2d})^{\mu\nu} + (\mbox{2e})^{\mu\nu} &=& \sqrt{|g|}T^{\mu\nu}|_{\Delta \phi,\mathcal{O}(\kappa^2)}, \nonumber \\
(\mbox{2f})^{\mu\nu} &=& \hat{t}^{\mu\nu}|_{\Delta\phi,\mathcal{O}(\kappa^2)}\equiv t^{\mu\nu}|_{\Delta\phi,\mathcal{O}(\kappa^2)},
\label{pseudoDilaton}
\end{eqnarray}
and the diagrams with internal gauge bosons represent
\begin{eqnarray}
(\mbox{2g})^{\mu\nu} + (\mbox{2h})^{\mu\nu} &=& \sqrt{|g|}T^{\mu\nu}|_{\Delta A,\mathcal{O}(\tilde{g}^2)}, \nonumber \\
(\mbox{2i})^{\mu\nu} &=& \hat{t}^{\mu\nu}|_{\Delta A,\mathcal{O}(\tilde{g}^2)}\equiv t^{\mu\nu}|_{\Delta A,\mathcal{O}(\tilde{g}^2)}.
\label{pseudoGauge}
\end{eqnarray}
Since $\left(1-\sqrt{|g|}\right)$ is purely gravitational, $\hat{t}^{\mu\nu}|_{\Delta \phi} \equiv t^{\mu\nu}|_{\Delta\phi}$ and $\hat{t}^{\mu\nu}|_{\Delta A} \equiv t^{\mu\nu}|_{\Delta A}$. Similar to Eq.~\eqref{matterRad}, the diagrams (2a), (2b), (2d), (2e), (2g), and (2h) sum to give $\sqrt{|g|}T^{\mu\nu}$, (2c), (2f), and (2i) give $\hat{t}^{\mu\nu}$, and all nine sum to give the locally conserved pseudotensor $\hat{T}^{\mu\nu}$, where all expressions are only kept to second order in $\kappa$ or $\tilde{g}$. While the diagrammatic representation may be useful for organizing higher order computations, it is simple enough to calculate $\sqrt{|g|}T^{\mu\nu}$ as a single algebraic expression for the purpose of confirming the validity of the radiative double copy to leading order.

Focusing on the energy-momentum tensor $\sqrt{|g|}T^{\mu\nu}$,
\begin{equation}
\sqrt{|g|}T^{\mu\nu} = \sum_{\alpha = 1}^N m_\alpha \int d\tau e^{ik\cdot x_\alpha(\tau)}\left(v_\alpha^\mu + \frac{dz_\alpha^\mu(\tau)}{d\tau} \right)\left(v_\alpha^\nu + \frac{dz_\alpha^\nu(\tau)}{d\tau}\right).
\end{equation}
Solving for the appropriate field equations gives $h^{\mu\nu}$, $\phi$, and $A^{\mu a}$ to lowest order,
\begin{eqnarray}
h^{\mu\nu}(x)|_{\mathcal{O}(\kappa^1)} &=& \frac{\kappa}{2} \sum_{\alpha = 1}^N m_\alpha\int_l (2\pi)\delta(l \cdot v_\alpha)\frac{e^{-il\cdot (x-b_\alpha)}}{l^2}\left(v_\alpha^\mu v_\alpha^\nu -\frac{\eta^{\mu\nu}}{d-2}\right), \nonumber \\
\phi(x)|_{\mathcal{O}(\kappa^2)} &=& \frac{1}{(d-2)}\left(\frac{\kappa}{2}\right)^2\sum_{\alpha = 1}^N m_\alpha \int_l (2\pi)\delta(l\cdot v_\alpha)\frac{e^{-il\cdot (x-b_\alpha)}}{l^2}, \nonumber \\
A^{\mu\tilde{a}}(x)|_{\mathcal{O}(\tilde{g}^1)} &=& -\tilde{g}\sum_{\alpha=1}^N \int_l(2\pi)\delta(l\cdot v_\alpha)\frac{e^{-il\cdot (x-b_\alpha)}}{l^2}c_\alpha^{\tilde{a}} v_\alpha^\mu .
\label{fieldSolutions}
\end{eqnarray}
Plugging the lowest-order field solutions into Eqs.~\eqref{GRdeflection},~\eqref{dilatonDeflection}, and \eqref{YMdeflection} gives
\begin{eqnarray}
\left.\frac{d^2 z_\alpha^\mu(\tau)}{d\tau^2}\right|_{\Delta h} &=& i\left(\frac{\kappa}{2}\right)^2\sum_{\beta\neq\alpha}m_\beta\int_{l_\beta}(2\pi)\delta(l_\beta\cdot v_\beta)\frac{e^{-il_\beta\cdot(x-b_\beta)}}{l_\beta^2}\nonumber \\
&& \null \times\left[2(v_\alpha\cdot v_\beta)k\cdot v_\alpha v_\beta^\mu - \frac{2k\cdot v_\alpha}{d-2}v_\alpha^\mu-\left((v_\alpha\cdot v_\beta)^2-\frac{1}{d-2}\right)l_\beta^\mu\right], \nonumber \\
\left.\frac{d^2 z_\alpha^\mu(\tau)}{d\tau^2}\right|_{\Delta \phi} &=&  \frac{-i}{d-2}\left(\frac{\kappa}{2}\right)^2\sum_{\beta\neq\alpha}m_\beta\int_{l_\beta}(2\pi)\delta(l_\beta\cdot v_\beta)\frac{e^{-il_\beta\cdot(x-b_\beta)}}{l_\beta^2} l_\beta^\mu, \\
\left.\frac{d^2z_\alpha^\mu}{d\tau^2}\right|_{\Delta A} &=& i\tilde{g}^2\sum_{\beta\neq\alpha}\frac{c_\alpha^{\tilde{a}}c_\beta^{\tilde{a}}}{m_\alpha}\int_l (2\pi)\delta(l\cdot v_\beta)\frac{e^{-il\cdot(b_{\alpha\beta} + v_\alpha\tau)}}{l^2}\left[(v_\alpha\cdot v_\beta) l^\mu - (l\cdot v_\alpha)v_\beta^\mu \right]. \nonumber
\end{eqnarray}

The corrections to the position are useful for finding $\sqrt{|g|}T^{\mu\nu}(k)$, 
\begin{eqnarray}
\sqrt{|g|}T^{\mu\nu}(k) &=& \sum_{\alpha = 1}^N m_\alpha \int d\tau e^{ik\cdot(b_\alpha + v_\alpha\tau + z_\alpha(\tau))}\left(v_\alpha^\mu + \frac{dz_\alpha^\mu(\tau)}{d\tau} \right)\left(v_\alpha^\nu + \frac{dz_\alpha^\nu(\tau)}{d\tau} \right), \nonumber \\
\sqrt{|g|}T^{\mu\nu}(k)|_{\mathcal{O}(\kappa^2)} &=& \sum_{\alpha = 1}^N m_\alpha \int d\tau e^{ik\cdot(b_\alpha + v_\alpha\tau)}\left[ ik\cdot z_\alpha v_\alpha^\mu v_\alpha^\nu + \frac{dz_\alpha^\mu}{d\tau}v_\alpha^\nu + v_\alpha^\mu \frac{dz_\alpha^\nu}{d\tau} \right].
\end{eqnarray}
The lowest order term proportional to $v_\alpha^\mu v_\alpha^\nu$ may be dropped, as it was used find the solution to $h^{\mu\nu}$ in Eq.~\eqref{fieldSolutions}. Focusing on the corrections due to gravity,
\begin{eqnarray}
\sqrt{|g|}T^{\mu\nu}(k)|_{\Delta h} &=& \left(\frac{\kappa}{2}\right)^2\sum_{\alpha\neq\beta}m_\alpha m_\beta \int_{l_\alpha,l_\beta}\mu_{\alpha,\beta}(k)l_\alpha^2 \\
&& \times\left[v_\alpha^\mu v_\alpha^\nu\left(2v_\alpha\cdot v_\beta \frac{k\cdot v_\beta}{k\cdot v_\alpha}+ \frac{2}{d-2}-\frac{l_\beta\cdot k}{(k\cdot v_\alpha)^2}\left((v_\alpha\cdot v_\beta)^2 - \frac{1}{d-2}\right)\right)\right.  \nonumber \\
&& \left.\null - 2v_\alpha\cdot v_\beta(v_\alpha^\mu v_\beta^\nu + v_\alpha^\nu v_\beta^\mu) + \frac{1}{k\cdot v_\alpha}\left((v_\alpha\cdot v_\beta)^2-\frac{1}{d-2}\right)(v_\alpha^\mu l_\beta^\nu + v_\alpha^\nu l_\beta^\mu)\right]. \nonumber
\end{eqnarray}
Additionally, the dilaton contributes
\begin{eqnarray}
\sqrt{|g|}T^{\mu\nu}(k)|_{\Delta\phi} &=& \frac{1}{d-2}\left(\frac{\kappa}{2}\right)^2\sum_{\alpha\neq\beta}m_\alpha m_\beta \int_{l_\alpha,l_\beta}\mu_{\alpha,\beta}(k)l_\alpha^2 \nonumber \\
&&\null \times\left[-v_\alpha^\mu v_\alpha^\nu \left(\frac{l_\beta\cdot k}{(k\cdot v_\alpha)^2}\right) + \frac{1}{k\cdot v_\alpha}(v_\alpha^\mu l_\beta^\nu + v_\alpha^\nu l_\beta^\mu)\right]. 
\end{eqnarray}
Finally, the gauge boson contributes
\begin{eqnarray}
\sqrt{|g|}T^{\mu\nu}(k)|_{\Delta A} &=& \tilde{g}^2\sum_{\substack{\alpha=1\\ \beta\neq\alpha}}^N c_\alpha^{\tilde{a}}c_\beta^{\tilde{a}}\int_{l_\alpha,l_\beta}\mu_{\alpha,\beta}(k)\frac{l_\alpha^2}{k\cdot v_\alpha}\left[ v_\alpha^\mu v_\alpha^\nu \left( v_\alpha\cdot v_\beta\frac{k\cdot l_\beta}{k\cdot v_\alpha} - k\cdot v_\beta\right) \right.\nonumber \\
&& \left. \null+(v_\alpha^\mu v_\beta^\nu + v_\alpha^\nu v_\beta^\mu)(k\cdot v_\alpha) - (v_\alpha^\mu l_\beta^\nu + v_\alpha^\nu l_\beta^\mu)(v_\alpha\cdot v_\beta)\right].
\end{eqnarray}

Appendix~\eqref{appendixB} works out the algebraic details of computing $t^{\mu\nu}|_{\Delta h}$ and $\hat{t}^{\mu\nu}|_{\Delta h}$. In summary, gravity contributes
\begin{eqnarray}
\label{2c}
\hat{t}^{\mu\nu}(k)|_{\Delta h} &=&  \left(\frac{\kappa}{2}\right)^2\sum_{\substack{\alpha=1\\ \beta\neq\alpha}}^N m_\alpha m_\beta\int_{l_\alpha,l_\beta} \mu_{\alpha,\beta}(k)\left[ 2v_\alpha^\mu v_\alpha^\nu\left((k\cdot v_\beta)^2-\frac{l_\alpha^2}{d-2}\right) \right. \\
&&\null+ \left(v_\alpha^\mu v_\beta^\nu + v_\alpha^\nu v_\beta^\mu\right) \left(l_\alpha^2 v_\alpha\cdot v_\beta - k\cdot v_\alpha k\cdot v_\beta\right)  -2 \left(v_\alpha^\mu l_\alpha^\nu + v_\alpha^\nu l_\alpha^\mu\right)\left(v_\alpha\cdot v_\beta k\cdot v_\beta\right) \nonumber \\
&& \left.\null+ l_\alpha^\mu l_\alpha^\mu\left(\!(v_\alpha\cdot v_\beta)^2-\frac{1}{d-2}\right)  + \eta^{\mu\nu}\left(\! k\cdot v_\alpha k\cdot v_\beta v_\alpha\cdot v_\beta -\frac{l_\alpha^2}{2}\!\left(\!(v_\alpha\cdot v_\beta)^2-\frac{1}{d-2}\!\right)\!\right) \right]. \nonumber 
\end{eqnarray}
Similarly, Eq.~\eqref{EYMdiagram} the dilaton contributes
\begin{equation}
\hat{t}^{\mu\nu}(k)|_{\Delta \phi} = \frac{1}{(d-2)}\left(\frac{\kappa}{2}\right)^2\sum_{\substack{\alpha=1\\ \beta\neq\alpha}}^N m_\alpha m_\beta\int_{l_\alpha,l_\beta} \mu_{\alpha,\beta}(k)\left[-l_\alpha^\mu l_\beta^\nu + \eta^{\mu\nu}\frac{l_\alpha\cdot l_\beta}{2} \right]. 
\end{equation}
When calculated algebraically from Eq.~\eqref{EYMdiagram}, the gauge boson contributes
\begin{eqnarray}
\label{EYMcontrib}
\hat{t}^{\mu\nu}(k)|_{\Delta A} &=& \tilde{g}^2 \sum_{\substack{\alpha=1\\ \beta\neq\alpha}}^N c_\alpha^{\tilde{a}}c_\beta^{\tilde{a}}\int_{l_\alpha,l_\beta}\mu_{\alpha,\beta}(k)\left[\frac{1}{2}\left(v_\alpha^\mu v_\beta^\nu+v_\alpha^\nu v_\alpha^\mu\right) l_\alpha\cdot l_\beta \right. \\
&& \left. \null+ \left(v_\alpha^\mu l_\alpha^\nu + v_\alpha^\nu l_\alpha^\mu\right)k\cdot v_\beta -l_\alpha^\mu l_\alpha^\nu v_\alpha\cdot v_\beta -\frac{1}{2}\eta^{\mu\nu}\left(k\cdot v_\alpha k\cdot v_\beta + v_\alpha\cdot v_\beta l_\alpha\cdot l_\beta\right)\right]. \nonumber
\end{eqnarray}
In Appendix \eqref{AppRadRules}, the three-point boson vertices of Einstein-Yang-Mills theory are used to find the same results for $t^{\mu\nu}$ via radiative Feynman rules.

Summing all contributions to order $\kappa^2$ gives the contributions from pure dilaton gravity,
\begin{eqnarray}
\hat{T}^{\mu\nu}(k)|_{\mathcal{O}(\kappa^2)} &=& \left(\frac{\kappa}{2}\right)^2\sum_{\substack{\alpha=1\\ \beta\neq\alpha}}^N m_\alpha m_\beta \int_{l_\alpha,l_\beta} \mu_{\alpha,\beta}(k)\left[ v_\alpha^\mu v_\alpha^\nu \left(2(k\cdot v_\beta)^2+2k\cdot v_\beta \frac{l_\alpha^2 v_\alpha\cdot v_\beta}{k\cdot v_\alpha} -\frac{l_\alpha^2 (v_\alpha\cdot v_\beta)^2k\cdot l_\beta}{(k\cdot v_\alpha)^2}\right)\right. \nonumber \\
&& \null - (v_\alpha^\mu v_\beta^\nu + v_\alpha^\nu v_\beta^\mu)\left(l_\alpha^2 v_\alpha\cdot v_\beta +k\cdot v_\alpha k\cdot v_\beta\right) - (v_\alpha^\mu l_\alpha^\nu + v_\alpha^\nu l_\alpha^\mu)(v_\alpha\cdot v_\beta)\left(\frac{l_\alpha^2 v_\alpha\cdot v_\beta}{k\cdot v_\alpha} + 2k\cdot v_\beta\right) \nonumber \\
&& \left.\null + l_\alpha^\mu l_\alpha^\nu (v_\alpha\cdot v_\beta)^2 + \eta^{\mu\nu}(v_\alpha\cdot v_\beta)\left(\frac{l_\alpha^2 v_\alpha \cdot v_\beta}{2}+k\cdot v_\alpha k\cdot v_\beta\right)\right].
\label{GRrad}
\end{eqnarray}
which agree with the results found in Ref.~\cite{Goldberger:2016iau}. $\sqrt{|g|}T^{\mu\nu}|_{\Delta A}$ and $\hat{t}^{\mu\nu}|_{\Delta A}$ give the additional contributions in Einstein-Yang-Mills theory,
\begin{eqnarray}
\hat{T}^{\mu\nu}(k)|_{\mathcal{O}(\tilde{g}^2)} &=& \tilde{g}^2\sum_{\substack{\alpha=1\\ \beta\neq\alpha}}^N\int_{l_\alpha,l_\beta}\mu_{\alpha,\beta}(k)\left[ v_\alpha^\mu v_\alpha^\nu\left(v_\alpha\cdot v_\beta\frac{k\cdot l_\beta}{(k\cdot v_\alpha)^2}-\frac{k\cdot v_\beta}{k\cdot v_\alpha}\right)l_\alpha^2 + \frac{1}{2}(v_\alpha^\mu v_\beta^\nu + v_\alpha^\nu v_\beta^\mu)l_\alpha^2\right.  \\
&& \left.\null +(v_\alpha^\mu l_\alpha^\nu + v_\alpha^\nu l_\alpha^\mu)\left(\frac{l_\alpha^2 v_\alpha\cdot v_\beta}{k\cdot v_\alpha} + k\cdot v_\beta\right) -l_\alpha^\mu l_\alpha^\nu v_\alpha\cdot v_\beta -\frac{1}{2}\eta^{\mu\nu}\left(k\cdot v_\alpha k\cdot v_\beta +l_\alpha^2 v_\alpha\cdot v_\beta\right) \right]. \nonumber 
\label{EYMradSource}
\end{eqnarray}
Adding this result to Eq.~\eqref{GRrad} gives the total source for gravitational radiation for Einstein-Yang-Mills theory. Next, we show that this result agrees precisely with what is found with the radiative double-copy method.

\subsection{The Radiative Double Copy}
In order to use the double copy to find gravitational radiation in Einstein-Yang-Mills theory, the same replacement rules used for general relativity~\cite{Goldberger:2016iau} may be used with the radiation found in Yang-Mills-biadjoint-scalar theory. The replacement rules are
\begin{eqnarray}
g &\rightarrow& \frac{\kappa}{2}, \nonumber \\
y &\rightarrow& \tilde{g}, \nonumber \\
c_\alpha^a &\rightarrow& p_\alpha^\nu, \nonumber \\
if^{a_1a_2a_3} &\rightarrow& -\frac{1}{2}(\eta^{\nu_1\nu_3}(q_1-q_3)^{\nu_2} + \eta^{\nu_1\nu_2}(q_2-q_1)^{\nu_3} + \eta^{\nu_2\nu_3}(q_3-q_2)^{\nu_1}), \nonumber \\
\hat{J}^{\mu a}(k) &\rightarrow& \hat{T}^{\mu\nu}(k),
\label{replacementRules}
\end{eqnarray}
where the momenta $q_1+q_2+q_3=0$. Similar to the Ward identity $k_\mu \hat{J}^{\mu a} = 0$, we can shift $\hat{T}^{\mu\nu}$ by terms proportional to either $k^\mu$ or $k^\nu$, such that $k_\mu \hat{T}^{\mu\nu} = k_\nu \hat{T}^{\mu\nu} = 0$, which shifts the gauge-dependent pseudotensor into the harmonic gauge. Since Ref.~\cite{Goldberger:2016iau} showed that the radiative double copy could recover $\hat{T}^{\mu\nu}|_{\mathcal{O}(\kappa^2)}$ and Ref.~\cite{Luna:2017dtq} showed how to use Yang-Mills ghosts to remove the dilaton, we focus on the additional terms introduced in Einstein-Yang-Mills theory. Applying the double copy replacement rules in Eq.~\eqref{replacementRules} to Eq.~\eqref{radBiadjoint} gives
\begin{eqnarray}
\hat{T}^{\mu\nu}(k)|_{\tilde{g}^2} &=& \tilde{g}^2\sum_{\substack{\alpha=1\\ \beta\neq\alpha}}^N m_\alpha m_\beta c_\alpha^{\tilde{a}}c_\beta^{\tilde{a}}\int_{l_\alpha,l_\beta}\mu_{\alpha,\beta}(k) \left[v_\alpha\cdot v_\beta\frac{l_\alpha^2v_\alpha^\nu}{k\cdot v_\alpha}\left(\frac{k\cdot l_\beta}{k\cdot v_\alpha}v_\alpha^\mu-l_\beta^\mu \right) \right. \\
&& \left.\null -\frac{1}{2}\left(2k\cdot v_\beta v_\alpha^\nu - 2k\cdot v_\alpha v_\beta^\nu + v_\alpha\cdot v_\beta (l_\beta - l_\alpha)^\nu\right)\left(\frac{l_\alpha^2}{k\cdot v_\alpha}v_\alpha^\mu - l_\alpha^\mu\right)\right]. \nonumber 
\end{eqnarray}
Shifting $l_\beta^\mu \rightarrow (l_\beta-l_\alpha)^\mu/2$ gives the gauge invariant $\hat{T}^{\mu\nu}$, 
\begin{eqnarray}
\hat{T}^{\mu\nu}(k)|_{\mathcal{O}(\tilde{g}^2)} &=& \tilde{g}^2\sum_{\substack{\alpha=1\\ \beta\neq\alpha}}^N m_\alpha m_\beta c_\alpha^{\tilde{a}}c_\beta^{\tilde{a}}\int_{l_\alpha,l_\beta}\mu_{\alpha,\beta}(k) \\
&&\times \left[v_\alpha\cdot v_\beta\frac{l_\alpha^2v_\alpha^\nu}{k\cdot v_\alpha}\left(\frac{k\cdot l_\beta}{k\cdot v_\alpha}v_\alpha^\mu-\frac{1}{2}(l_\beta-l_\alpha)^\mu \right) \right. \nonumber \\
&& \left. \null-\frac{1}{2}\left(2k\cdot v_\beta v_\alpha^\nu - 2k\cdot v_\alpha v_\beta^\nu + v_\alpha\cdot v_\beta (l_\beta - l_\alpha)^\nu\right)\left(\frac{l_\alpha^2}{k\cdot v_\alpha}v_\alpha^\mu - l_\alpha^\mu\right)\right]. \nonumber
\end{eqnarray}
Symmetrizing this result gives the appropriate final expression for $\hat{T}^{\mu\nu}$,
\begin{eqnarray}
\hat{T}^{\mu\nu}|_{\mathcal{O}(\tilde{g}^2)} &=& -\tilde{g}^2\left(\frac{\kappa}{2}\right)^2\sum_{\substack{\alpha=1\\ \beta\neq\alpha}}^N m_\alpha m_\beta c_\alpha^{\tilde{a}}c_\beta^{\tilde{a}} \int_{l_\alpha,l_\beta}\mu_{\alpha,\beta}(k) \\
&& \times\left[ v_\alpha^\mu v_\alpha^\nu\left(\frac{k\cdot v_\beta}{k\cdot v_\alpha} - \frac{v_\alpha\cdot v_\beta}{(k\cdot v_\alpha)^2}k\cdot l_\beta\right)l_\alpha^2-\frac{1}{2}(v_\alpha^\mu v_\beta^\nu + v_\alpha^\nu v_\beta^\mu)l_\alpha^2\right. \nonumber \\
&& \left.\null -(v_\alpha^\mu l_\alpha^\nu + v_\alpha^\nu l_\alpha^\mu)\left(\frac{v_\alpha\cdot v_\beta}{k\cdot v_\alpha}l_\alpha^2 +k\cdot v_\beta\right) + l_\alpha^\mu l_\alpha^\nu(v_\alpha\cdot v_\beta) + \frac{1}{2}\eta^{\mu\nu} l_\alpha^2(v_\alpha\cdot v_\beta) \right], \nonumber
\label{EYMradDouble}
\end{eqnarray}
where the gauge condition allows for $v_\alpha^\mu k^\nu = \frac{1}{2}\eta^{\mu\nu} k\cdot v_\alpha$. This result agrees precisely with what was found in Eq.~\eqref{EYMradSource}, demonstrating that the radiative double copy holds for Einstein-Yang-Mills theory to leading order.

\subsection{Einstein-Maxwell Theory}

Since it is more physically relevant to scatter massive point particles with electric charge rather than particles with weak-isospin or color, an Abelian $U(1)$ gauge symmetry is also worth studying. The action for fields in Einstein-Maxwell theory is
\begin{equation}
S = \int d^dx \sqrt{|g|}\left(-\frac{2}{\kappa^2} R - \frac{1}{4}g^{\mu\rho}g^{\nu\sigma} F_{\mu\nu}F_{\rho\sigma}\right) .
\end{equation}
When comparing with Einstein-Yang-Mills theory, the Maxwell field $A^{\mu}$ can be recovered from a single component of the Yang-Mills field $A^{\mu\tilde{a}}$. In order to find results in Einstein-Maxwell theory from Einstein-Yang-Mills theory, care must be taken with the coupling constants. For example, the Maxwell current density for point particles is given by
\begin{equation}
J^{\mu}(x) = e\sum_{\alpha = 1}^N q_\alpha \int d\tau v_\alpha^{\mu}(\tau)\delta^d(x-x_\alpha(\tau)),
\end{equation}
where $q_\alpha = -1$ for electrons, such that $eq_\alpha$ represents the electric charge of particle $\alpha$. In order to recover Einstein-Maxwell theory from Einstein-Yang-Mills, one must substitute $\tilde{g}\rightarrow e$ and $c_\alpha^{\tilde{a}} \rightarrow q_\alpha$, given our conventions for $\tilde{g}$ and the normalization of the Lagrangian given in Eq.~\eqref{gravAction}. Applying these substitutions to Eq.~\eqref{EYMradSource} would give gravitational radiation in Einstein-Maxwell theory. At higher orders, $f^{\tilde{a}\tilde{b}\tilde{c}}$ would be sent to zero as well.

In terms of the radiative double copy, an adjoint scalar field $\Phi^a$ could also be seen as a single component of the biadjoint scalar feld $\Phi^{\tilde{a}a}$. Results for Yang-Mills-adjoint-scalar theory can easily be found from Eq.~\eqref{radBiadjoint} by properly sending $c_\alpha^{\tilde{a}}\rightarrow q_\alpha$ and reinterpreting $y$ as the coupling constant of the adjoint scalar theory. It is straightforward to see that the double copy of Yang-Mills-adjoint-scalar theory gives solutions in Einstein-Maxwell theory with the replacement rules shown in Eq.~\eqref{replacementRules} and $y\rightarrow e$.

\section{Conclusions}
\label{sec5}

In previous work, the double copy has been applied to gravitational radiation in general relativity with a dilaton, which suggested that schematic radiative diagrams may be useful for depicting sources of radiation~\cite{Goldberger:2016iau}. Similarly, it was shown that the same replacement rules can be used to find Yang-Mills radiation from biadjoint-scalar radiation~\cite{Goldberger:2017frp} and that ghosts can be used to remove the dilaton~\cite{Luna:2017dtq}.

In this work, the gravitational radiation produced by colliding color charges was found within the context of Einstein-Yang-Mills theory. Our primary result demonstrates that the double copy can be used to find radiation in Einstein-Yang-Mills theory from Yang-Mills-biadjoint-scalar theory. These calculations provided insight on how a radiative diagrammatic scheme closer to Feynman diagrams used for scattering amplitudes may be possible. Furthermore, radiation in Einstein-Maxwell theory can be found via similar methods. This work suggests that it may be possible to develop systematic rules for constructing radiative diagrams that can be used to calculate radiation to higher orders, at least for initial conditions associated with $N$ particle scattering. It appears that rules for worldline propagators would be needed, in addition to the typical rules used for scattering amplitudes. 

In future work, it would be interesting to investigate if the radiative double copy holds for higher orders, as the precise replacement rules are not yet known. Additional efforts to perform the integrals are also needed. The gravitational interactions between the quantized spin of Dirac particles would be an interesting theoretical challenge, while considering the scattering of macroscopic mass distributions with classical angular momentum would be more applicable for experiments such as LIGO. Studying the formation of bound states due to higher order effects would also be important.

The author would like to thank Zvi Bern for constant guidance, as well as Walter Goldberger, Donal O'Connell, Alexander Ridgway, Jedidiah Thompson, and Julio Parra-Martinez for various discussions.

\appendix

\section{Derivation of Gravitational Radiation from Pseudotensor}
\label{appendixB}

In this section, the steps for deriving the gravitational radiation coming from nonlinear gravitational interactions are provided. In Section~\eqref{sec4}, Einstein's field equations to first order for weak gravitational fields was found to be
\begin{equation}
\Box\bar{h}^{\mu\nu} = -\frac{\kappa}{2}\hat{T}^{\mu\nu},
\end{equation}
where the energy-momentum pseudotensor $\hat{T}^{\mu\nu} = T^{\mu\nu} + t^{\mu\nu} = \sqrt{|g|}T^{\mu\nu} + \hat{t}^{\mu\nu}$ contains the nonlinear corrections to the linearized field equations, such that the purely gravitational component of the pseudotensor $t^{\mu\nu}$ is given by Eq.~\eqref{EYMdiagram}
\begin{eqnarray}
\label{pseudoAppB}
t^{\mu\nu} &=& 2h_{\rho\sigma}\left(h^{\mu\rho,\nu\sigma} + h^{\nu\sigma,\mu\rho} - h^{\mu\nu,\rho\sigma} - h^{\rho\sigma,\mu\nu} \right) + h^{\mu\nu}\Box h - 2h^{\mu\rho}\Box h_\rho^\nu - 2h^{\nu\rho}\Box h_\rho^\mu  \\
&& \null-2h^{\mu\rho,\sigma}\left(h^\nu_{\rho,\sigma} - h^\nu_{\sigma,\rho}\right) - h^{\rho\sigma,\mu}h_{\rho\sigma}^{\textrm{ },\nu} + \eta^{\mu\nu}\left[2h^{\rho\sigma}\Box h_{\rho\sigma} + h_{\rho\sigma,\lambda}\left(\frac{3}{2}h^{\rho\sigma,\lambda} - h^{\rho\lambda,\sigma}\right)\right]. \nonumber
\end{eqnarray}
In order to solve for this, the lowest-order solution of the gravitational field is used
\begin{equation}
h^{\mu\nu}(x) = \frac{\kappa}{2}\sum_{\alpha = 1}^N m_\alpha \int_{l_\alpha}(2\pi)\delta(l_\alpha\cdot v_\alpha) \frac{e^{-il_\alpha\cdot(x-b_\alpha)}}{l_\alpha^2}\left(v_\alpha^\mu v_\alpha^\nu - \frac{\eta^{\mu\nu}}{d-2} \right),
\end{equation}
which gives rise to a source for the nonlinear gravitational interaction via $t^{\mu\nu}$. Each term in $t^{\mu\nu}$ is second order in $h^{\mu\nu}$, so one is related to particle $\alpha$ and another to particle $\beta$, giving a double sum. The summation and integrals on all terms will have the following form
\begin{equation}
t^{\mu\nu} =\left(\frac{\kappa}{2}\right)^2 \sum_{\substack{\alpha=1\\ \beta\neq\alpha}}^N m_\alpha m_\beta \int_{l_\alpha,l_\beta}\mu_{\alpha,\beta}(k) I^{\mu\nu},
\end{equation}
where $I^{\mu\nu}$ is the integrand containing many terms. For the integrand, focusing on the $\left(v_\alpha^\mu v_\alpha^\nu - \eta^{\mu\nu}/(d-2)\right)$ portion of the solution to $h^{\mu\nu}$ and manually plug these pieces into Eq.~\eqref{pseudoAppB} gives
\begin{eqnarray}
I^{\mu\nu} &=& 2\left(v_{\beta\rho}v_{\beta\sigma} - \frac{\eta_{\rho\sigma}}{d-2}\right)\left[-l_\alpha^\nu l_\alpha^\sigma\left(v_\alpha^\mu v_\alpha^\rho - \frac{\eta^{\mu\rho}}{d-2}\right) - l_\alpha^\mu l_\alpha^\rho\left(v_\alpha^\nu v_\alpha^\sigma - \frac{\eta^{\nu\sigma}}{d-2}\right)  \right. \nonumber \\
&&\left. \mbox{ }\mbox{ }\mbox{ }\mbox{ }\mbox{ }\null+ l_\alpha^\rho l_\alpha^\sigma\left(v_\alpha^\mu v_\alpha^\nu - \frac{\eta^{\mu\nu}}{d-2}\right) + l_\alpha^\mu l_\alpha^\nu\left(v_\alpha^\rho v_\alpha^\sigma - \frac{\eta^{\rho\sigma}}{d-2}\right) \right] -l_\beta^2\left(v_\alpha^\mu v_\alpha^\nu - \frac{\eta^{\mu\nu}}{d-2}\right)\left(\frac{-2}{d-2}\right)\nonumber \\
&& \null+2l_\alpha^2\left(v_\beta^\mu v_\beta^\rho - \frac{\eta^{\mu\rho}}{d-2}\right)\left(v_{\alpha\rho} v_\alpha^\nu - \frac{\eta_\rho^{\nu}}{d-2}\right)+2l_\alpha^2\left(v_\beta^\nu v_\beta^\rho - \frac{\eta^{\nu\rho}}{d-2}\right)\left(v_{\alpha\rho} v_\alpha^\mu - \frac{\eta_\rho^{\mu}}{d-2}\right) \nonumber \\
&& \null -2il_\alpha^\sigma\left(v_\alpha^\mu v_\alpha^\rho - \frac{\eta^{\mu\rho}}{d-2}\right)\left[il_{\beta\sigma}\left(v_\beta^\nu v_{\beta\rho} - \frac{\eta^\nu_\rho}{d-2}\right)-il_{\beta\rho}\left(v_\beta^\nu v_{\beta\sigma} - \frac{\eta^\nu_\sigma}{d-2}\right)\right] \nonumber \\
&& \null+ l_\alpha^\mu l_\beta^\nu\left(v_\alpha^\rho v_\alpha^\sigma - \frac{\eta^{\rho\sigma}}{d-2}\right)\left(v_{\beta\rho} v_{\beta\sigma} - \frac{\eta_{\rho\sigma}}{d-2}\right)\nonumber \\
&& \null+ \eta^{\mu\nu}\left\{ -2l_\alpha^2\left(v_\alpha^\rho v_\alpha^\sigma - \frac{\eta^{\rho\sigma}}{d-2}\right)\left(v_{\beta\rho} v_{\beta\sigma} - \frac{\eta_{\rho\sigma}}{d-2}\right)\right. \nonumber \\
&& \left. \null+il_{\alpha\lambda}\left(v_{\alpha\rho} v_{\alpha\sigma} - \frac{\eta_{\rho\sigma}}{d-2}\right)\left[\frac{3}{2}il_\beta^\lambda\left(v_\beta^\rho v_\beta^\sigma - \frac{\eta^{\rho\sigma}}{d-2}\right) - il_\beta^\sigma\left(v_\beta^\rho v_\beta^\lambda - \frac{\eta^{\rho\lambda}}{d-2}\right)\right]\right\}.
\end{eqnarray}

Distributing these factors and reorganizing all of the terms with the same tensor index structure gives
\begin{eqnarray}
I^{\mu\nu} &=& v_\alpha^\mu v_\alpha^\nu \left(2(k\cdot v_\beta)^2-\frac{2l_\alpha^2}{d-2} + \frac{2l_\beta^2}{d-2} - \frac{4l_\alpha^2 + 4l_\beta^2}{d-2} - \frac{4l_\alpha\cdot l_\beta}{d-2}\right) \nonumber \\
&& \null+ \left(v_\alpha^\mu v_\beta^\nu + v_\alpha^\nu v_\beta^\mu\right)\left(2l_\alpha^2(v_\alpha\cdot v_\beta) + l_\alpha\cdot l_\beta(v_\alpha\cdot v_\beta)-k\cdot v_\alpha k\cdot v_\beta\right) \nonumber \\
&& \null+ (v_\alpha^\mu l_\alpha^\nu + v_\alpha^\nu l_\alpha^\mu)\left(-2(v_\alpha\cdot v_\beta)k\cdot v_\beta+\frac{2 k\cdot v_\alpha}{d-2} \right) + (v_\alpha^\mu l_\beta^\nu + v_\alpha^\nu l_\beta^\mu)\left(\frac{2k\cdot v_\alpha}{d-2}\right) \nonumber \\
&& \null+ l_\alpha^\mu l_\alpha^\nu \left(-\frac{4}{(d-2)^2}+ 2(v_\alpha\cdot v_\beta)^2-\frac{4}{d-2}+\frac{2d}{(d-2)^2} \right) \nonumber \\
&& \null+ (l_\alpha^\mu l_\beta^\nu + l_\alpha^\nu l_\beta^\mu)\left(-\frac{1}{(d-2)^2}+\frac{1}{2}(v_\alpha\cdot v_\beta)^2 -\frac{1}{d-2}+\frac{d}{2(d-2)^2}\right) \nonumber \\
&& \null+ \eta^{\mu\nu}\left[-\frac{2(k\cdot v_\beta)^2}{d-2} + \frac{2l_\alpha^2}{(d-2)^2}-\frac{2l_\beta^2}{(d-2)^2}+\frac{4l_\alpha^2}{(d-2)^2}+\frac{2l_\alpha\cdot l_\beta}{(d-2)^2}\right.  \\
&& \mbox{ }\mbox{ }\mbox{ }\mbox{ }\mbox{ } \null-\left(2l_\alpha^2+\frac{3}{2}l_\alpha\cdot l_\beta\right)\left((v_\alpha\cdot v_\beta)^2-\frac{2}{d-2}+\frac{d}{(d-2)^2}\right) \nonumber \\
&&\left.\mbox{ }\mbox{ }\mbox{ }\mbox{ }\mbox{ }\null+ \left(v_\alpha\cdot v_\beta k\cdot v_\alpha k\cdot v_\beta + \frac{l_\alpha\cdot l_\beta}{(d-2)^2}\right)\right]. \nonumber
\end{eqnarray}
Next, the relation $k^2 = l_\alpha^2 + 2l_\alpha\cdot l_\beta + l_\beta^2 = 0$ is used to simplify further. The identity $a^\mu l_\beta^\nu  = a^\mu k^\nu - a^\mu l_\alpha^\nu$ and the gauge condition of the gravitational field allows for the gauge-invariant shift $a^\mu l_\beta^\nu \rightarrow \frac{1}{2} a\cdot k\eta^{\mu\nu} - a^\mu l_\alpha^\nu$, since dotting this expression with the polarization tensor would give the same radiation amplitude. Making such changes gives
\begin{eqnarray}
I^{\mu\nu} &=&  v_\alpha^\mu v_\alpha^\nu \left(2(k\cdot v_\beta)^2 -\frac{4l_\alpha^2}{d-2}\right) + (v_\alpha^\mu v_\beta^\nu + v_\alpha^\nu v_\beta^\mu) \left(l_\alpha^2(v_\alpha\cdot v_\beta) -k\cdot v_\alpha k\cdot v_\beta\right) \nonumber \\
&& \null-2(v_\alpha\cdot v_\beta)k\cdot v_\beta (v_\alpha^\mu l_\alpha^\nu + v_\alpha^\nu l_\alpha^\mu) + l_\alpha^\mu l_\alpha^\nu\left((v_\alpha\cdot v_\beta)^2-\frac{1}{d-2}\right) \nonumber \\
&& \null+ \eta^{\mu\nu} \left(\frac{2(k\cdot v_\alpha)^2}{d-2} +\frac{k\cdot l_\alpha}{2}(v_\alpha\cdot v_\beta)^2 - \frac{k\cdot l_\alpha}{2(d-2)}\right) \nonumber \\
&& \null+ \eta^{\mu\nu}\left[-\frac{2(k\cdot v_\beta)^2}{d-2} + \frac{2l_\alpha^2}{(d-2)^2}-\frac{2l_\beta^2}{(d-2)^2}+\frac{4l_\alpha^2}{(d-2)^2}+\frac{2l_\alpha\cdot l_\beta}{(d-2)^2}\right.  \\
&& \mbox{ }\mbox{ }\mbox{ }\mbox{ }\mbox{ }\null -\left(2l_\alpha^2+\frac{3}{2}l_\alpha\cdot l_\beta\right)\left((v_\alpha\cdot v_\beta)^2-\frac{2}{d-2}+\frac{d}{(d-2)^2}\right)\nonumber \\
&& \left. \mbox{ }\mbox{ }\mbox{ }\mbox{ }\mbox{ }\null+  \left(v_\alpha\cdot v_\beta k\cdot v_\alpha k\cdot v_\beta + \frac{l_\alpha\cdot l_\beta}{(d-2)^2}\right)\right]. \nonumber
\end{eqnarray}
By considering that $\alpha$ and $\beta$ are symmetric, all particle labels may be switched for any term, which allows further simplification to give the final result
\begin{eqnarray}
I^{\mu\nu} &=&  v_\alpha^\mu v_\alpha^\nu \left(2(k\cdot v_\beta)^2 -\frac{4l_\alpha^2}{d-2}\right) + (v_\alpha^\mu v_\beta^\nu + v_\alpha^\nu v_\beta^\mu) \left(l_\alpha^2(v_\alpha\cdot v_\beta) -k\cdot v_\alpha k\cdot v_\beta\right) \nonumber \\
&& \mbox{ }\mbox{ }\mbox{ }\mbox{ }\mbox{ }\null-2(v_\alpha\cdot v_\beta)k\cdot v_\beta (v_\alpha^\mu l_\alpha^\nu + v_\alpha^\nu l_\alpha^\mu) + l_\alpha^\mu l_\alpha^\nu\left((v_\alpha\cdot v_\beta)^2-\frac{1}{d-2}\right) \nonumber \\
&& \mbox{ }\mbox{ }\mbox{ }\mbox{ }\mbox{ }\null+ \eta^{\mu\nu} \left(v_\alpha\cdot v_\beta k\cdot v_\alpha k\cdot v_\beta - \frac{l_\alpha^2}{2}\left( (v_\alpha\cdot v_\beta)^2 - \frac{1}{d-2}\right)\right).
\end{eqnarray}
To more easily compare with the diagrammatic method, $\hat{t}^{\mu\nu}$ is found by adding the lowest-order term of $\left(1-\sqrt{|g|}\right)T^{\mu\nu}$, where
\begin{eqnarray}
T^{\mu\nu}(x) &\approx& \sum_{\alpha=1}^N m_\alpha \int_{l_\alpha}(2\pi) \delta(v_\alpha\cdot l_\alpha)e^{-il_\alpha\cdot (x-b_\alpha)}v_\alpha^\mu v_\alpha^\nu, \nonumber \\
h(x) &\approx& \frac{-\kappa}{d-2}\sum_{\beta\neq\alpha}m_\beta \int_{l_\beta}(2\pi)\delta(l_\beta \cdot v_\beta)\frac{e^{-il_\beta\cdot (x-b_\beta)}}{l_\beta^2},\nonumber \\
\left(1-\sqrt{|g|}\right)T^{\mu\nu} &\approx& \frac{1}{d-2}\left(\frac{\kappa}{2}\right)^2\sum m_\alpha m_\beta\int_{l_\alpha,l_\beta} \mu_{\alpha,\beta}(k) 2l_\alpha^2 v_\alpha^\mu v_\alpha^\nu.
\end{eqnarray}
Adding this to $t^{\mu\nu}$ gives $\hat{t}^{\mu\nu} \equiv \left(\frac{\kappa}{2}\right)^2 \sum_{\substack{\alpha=1\\ \beta\neq\alpha}}^N m_\alpha m_\beta \int_{l_\alpha,l_\beta}\mu_{\alpha,\beta}(k)\hat{I}^{\mu\nu}$, such that
\begin{eqnarray}
\hat{I}^{\mu\nu} &=& v_\alpha^\mu v_\alpha^\nu \left(2(k\cdot v_\beta)^2 -\frac{2l_\alpha^2}{d-2}\right) + (v_\alpha^\mu v_\beta^\nu + v_\alpha^\nu v_\beta^\mu) \left(l_\alpha^2(v_\alpha\cdot v_\beta) -k\cdot v_\alpha k\cdot v_\beta\right) \nonumber \\
&&\null -2(v_\alpha\cdot v_\beta)k\cdot v_\beta (v_\alpha^\mu l_\alpha^\nu + v_\alpha^\nu l_\alpha^\mu) + l_\alpha^\mu l_\alpha^\nu\left((v_\alpha\cdot v_\beta)^2-\frac{1}{d-2}\right) \nonumber \\
&&\null + \eta^{\mu\nu} \left(v_\alpha\cdot v_\beta k\cdot v_\alpha k\cdot v_\beta - \frac{l_\alpha^2}{2}\left( (v_\alpha\cdot v_\beta)^2 - \frac{1}{d-2}\right)\right).
\end{eqnarray}
As shown in the next appendix, this result agrees precisely with a diagram involving the three-point graviton vertex.

\section{Some Radiative Feynman Rules}
\label{AppRadRules}

\subsection{Yang-Mills and Biadjoint-Scalar Theory}

A Feynman diagram approach can be used to find the results for diagrams (1c) and (1f), shown in Fig.~\eqref{figYMbiadj}. Expanding the kinetic term of the Lagrangian, the $\mathcal{O}(A^3)$ term corresponding to the three-point vector boson interaction is
\begin{equation}
-\frac{1}{4} F_{\mu\nu}^a F^{\mu\nu a} = -\partial_\mu A_\nu^a g f^{abc} A^{\mu b} A^{\nu c} + \dots .
\end{equation}
This term in the Lagrangian gives the textbook non-Abelian three-point vector boson vertex, given by
\begin{equation}
\Gamma^{\mu a,\nu b,\rho c}(k,p,q) = f^{abc}\left((k^\nu - q^\nu) \eta^{\mu\rho} +  (p^\rho - k^\rho) \eta^{\nu\mu} +  (q^\mu - p^\mu) \eta^{\rho\nu}\right),
\end{equation}
where $A_\mu^a$ is associated with the momentum $k$, $A_\nu^b$ is associated with $p$, and $A_\rho^c$ is associated with $q$. 

The three-point vertex for two biadjoint scalars and one adjoint vector field can be used to efficiently calculate a piece radiation, which comes from the kinetic term of the biadjoint scalar. Focusing on the terms in the Lagrangian to $\mathcal{O}(\Phi^2 A)$,
\begin{equation}
\frac{1}{2}(D_\mu \Phi^{\tilde{a}})^a(D^\mu \Phi^{\tilde{b}})^a \delta^{\tilde{a}\tilde{b}} = g f^{abc} \delta^{\tilde{a}\tilde{c}} (\partial_\mu \Phi^{\tilde{a}a})A^{\mu b}\Phi^{\tilde{c}c} + \dots .
\end{equation}
Taking the appropriate functional derivatives and properly symmetrizing gives the three-point vertex for two scalars and one vector,
\begin{equation}
\Gamma^{\tilde{a}a,\nu b,\tilde{c}c}(k,p,q) = f^{abc} \delta^{\tilde{a}\tilde{c}} \left(k^\nu - q^\nu \right).
\end{equation}

The three-point vertices above can be used to find diagrams (1c) and (1f), giving
\begin{eqnarray}
(\mbox{1c})^{\mu a}(k) = \frac{1}{2}\int_{l_\alpha,l_\beta}A_\nu^{b}(l_\alpha)|_{\mathcal{O}(g^1)} i\Gamma^{\mu a,\nu b,\rho c}(-k,l_\alpha,l_\beta)A_\rho^{c}(l_\beta)|_{\mathcal{O}(g^1)}(2\pi)^d\delta^d(k-l_\alpha-l_\beta),\nonumber \\
(\mbox{1f})^{\mu a}(k) = \frac{1}{2}\int_{l_\alpha,l_\beta}\Phi^{\tilde{b}b}(l_\alpha)|_{\mathcal{O}(y^1)} i\Gamma^{\tilde{b}b,\mu a,\tilde{c}c}(l_\alpha,-k,l_\beta)\Phi^{\tilde{c}c}(l_\beta)|_{\mathcal{O}(y^1)}(2\pi)^d\delta^d(k-l_\alpha-l_\beta),
\end{eqnarray}
where a symmetry factor of $1/2$ has been added.

The solutions needed for these diagrams were found in Eq.~\eqref{BiadjointFieldSolution}, giving
\begin{eqnarray}
A^{\mu a}(l_\alpha)|_{\mathcal{O}(g^1)} &=& -g\sum_{\alpha = 1}^N (2\pi)\delta(l_\alpha \cdot v_\alpha)\frac{e^{il_\alpha\cdot b_\alpha}}{l_\alpha^2} v_\alpha^\mu c_\alpha^a, \nonumber \\
\Phi^{a \tilde{a}}(l_\alpha)|_{\mathcal{O}(y^1)} &=& -y \sum_{\alpha = 1}^N (2\pi)\delta(l_\alpha\cdot v_\alpha)\frac{e^{il_\alpha\cdot b_\alpha}}{l_\alpha^2}c_\alpha^a c_\alpha^{\tilde{a}}.
\end{eqnarray}
Plugging in these solutions gives
\begin{eqnarray}
(\mbox{1c})^{\mu a}(k) &=& \frac{g^2}{2}\sum_{\alpha\neq\beta}if^{abc}c_\alpha^b c_\beta^c \int_{l_\alpha,l_\beta}\mu_{\alpha,\beta}(k) \left[-2k\cdot v_\alpha v_\beta^\mu +2k\cdot v_\beta v_\alpha^\mu +v_\alpha\cdot v_\beta(l_\beta -l_\alpha)^\mu\right],\nonumber \\
(\mbox{1f})^{\mu a}(k) &=& \frac{y^2}{2}\sum_{\beta\neq\alpha}if^{abc}c_\alpha^b c_\beta^c c_\alpha^{\tilde{a}}c_\beta^{\tilde{a}}\int_{l_\alpha,l_\beta}\mu_{\alpha,\beta}(k)(l_\alpha - l_\beta)^\mu.
\end{eqnarray}
Due to the antisymmetry of $f^{abc}c_\alpha^b c_\beta^c$, switching $\alpha \leftrightarrow\beta$ for a term multiplied by this factor introduces a minus sign, allowing further simplification,
\begin{eqnarray}
(\mbox{1c})^{\mu a}(k) &=& g^2\sum_{\alpha\neq\beta}if^{abc}c_\alpha^b c_\beta^c \int_{l_\alpha,l_\beta}\mu_{\alpha,\beta}(k) \left[2k\cdot v_\beta v_\alpha^\mu -(v_\alpha\cdot v_\beta)l_\alpha^\mu\right],\nonumber \\
(\mbox{1f})^{\mu a}(k) &=& y^2\sum_{\beta\neq\alpha}if^{abc}c_\alpha^b c_\beta^c c_\alpha^{\tilde{a}}c_\beta^{\tilde{a}}\int_{l_\alpha,l_\beta}\mu_{\alpha,\beta}(k)l_\alpha^\mu.
\end{eqnarray}
Note how this result agrees with the algebraic method found in Eq.~\eqref{diag1f}.

\subsection{General Relativity and Einstein-Yang-Mills Theory}
Next, the three-point graviton vertex will be used to stitch together lower order gravitational field solutions to generate a piece of the gravitational radiation field. The three-point graviton vertex from DeWitt~\cite{DeWitt} and utilized by Sannan~\cite{Sannan} is
\begin{eqnarray}
V_{\mu\alpha,\nu\beta,\sigma\gamma}(k_1,k_2,k_3) &=& \textrm{sym}\left[ -\frac{1}{2}P_3(k_1\cdot k_2 \eta_{\mu\alpha}\eta_{\nu\beta}\eta_{\sigma\gamma}) -\frac{1}{2}P_6(k_{1\nu}k_{1\beta}\eta_{\mu\alpha}\eta_{\sigma\gamma})\right. \\
&& \null+ \frac{1}{2} P_3(k_1\cdot k_2 \eta_{\mu\nu}\eta_{\alpha\beta}\eta_{\sigma\gamma}) + P_6(k_1\cdot k_2\eta_{\mu\alpha}\eta_{\nu\sigma}\eta_{\beta\gamma}) + 2P_3(k_{1\nu}k_{1\gamma}\eta_{\mu\alpha}\eta_{\beta\sigma}) \nonumber \\
&& \null- P_3(k_{1\beta}k_{2\mu}\eta_{\alpha\nu}\eta_{\sigma\gamma}) +  P_3(k_{1\sigma}k_{2\gamma}\eta_{\mu\nu}\eta_{\alpha\beta}) + P_6(k_{1\sigma}k_{1\gamma}\eta_{\mu\nu}\eta_{\alpha\beta})\nonumber \\
&& \left. \null+ 2 P_6(k_{1\nu}k_{2\gamma}\eta_{\beta\mu}\eta_{\alpha\sigma})+ 2P_3(k_{1\nu}k_{2\mu}\eta_{\beta\sigma}\eta_{\gamma\alpha}) - 2P_3(k_1\cdot k_2 \eta_{\alpha\nu}\eta_{\beta\sigma}\eta_{\gamma\mu})\frac{}{}\right], \nonumber
\end{eqnarray}
where $P_3$ and $P_6$ refers to a permutation of $k_1$, $k_2$, and $k_3$ resulting in 3 or 6 terms, respectively, and $\textrm{sym}$ applies a symmetrization across $\mu\alpha$, $\nu\beta$, and $\sigma\gamma$. For example,
\begin{eqnarray}
P_3(k_1\cdot k_2\eta_{\mu\nu}\eta_{\alpha\beta}\eta_{\sigma\gamma}) &=& k_1\cdot k_2 \eta_{\mu\nu}\eta_{\alpha\beta}\eta_{\sigma\gamma} + k_2\cdot k_3\eta_{\nu\sigma}\eta_{\beta\gamma}\eta_{\mu\alpha}+ k_3\cdot k_1\eta_{\mu\sigma}\eta_{\alpha\gamma}\eta_{\nu\beta}, \nonumber \\
\textrm{sym}[\eta_{\mu\nu}\eta_{\alpha\beta}] &=& \frac{1}{4}\left(\eta_{\mu\nu}\eta_{\alpha\beta}+\eta_{\mu\beta}\eta_{\nu\alpha} + \eta_{\nu\alpha}\eta_{\mu\beta} + \eta_{\alpha\beta}\eta_{\mu\nu}\right).
\end{eqnarray}
Expanding $P_3$ and $P_6$ gives
\begin{eqnarray}
V^{\mu\alpha,\nu\beta,\sigma\gamma}(k_1,k_2,k_3) &=&\textrm{sym}\left[ -\frac{1}{2}\left(k_1\cdot k_2+k_2\cdot k_3 + k_3 \cdot k_1\right) \eta^{\mu\alpha}\eta^{\nu\beta}\eta^{\sigma\gamma} \right.\nonumber \\
&&\null -\frac{1}{2}\left(k_1^\nu k_1^\beta\eta^{\mu\alpha}\eta^{\sigma\gamma}+k_1^\sigma k_1^\gamma\eta^{\mu\alpha}\eta^{\nu\beta} k_2^\mu k_2^\alpha\eta^{\nu\beta}\eta^{\sigma\gamma} \right. \nonumber \\
&& \left.\null+ k_2^\sigma k_2^\gamma \eta^{\mu\alpha}\eta^{\nu\beta} + k_3^\mu k_3^\alpha\eta^{\nu\beta}\eta^{\gamma\sigma} + k_3^\nu k_3^\beta \eta^{\mu\alpha}\eta^{\gamma\sigma}\right) \nonumber \\
&& \null+ \frac{1}{2}\left(k_1\cdot k_2 \eta^{\mu\nu}\eta^{\alpha\beta}\eta^{\sigma\gamma} + k_2\cdot k_3 \eta^{\nu\sigma}\eta^{\beta\gamma}\eta^{\mu\alpha} + k_3\cdot k_1 \eta^{\mu\sigma}\eta^{\alpha\gamma}\eta^{\nu\beta}\right) \nonumber \\
&&\null+ \left(k_1\cdot k_2\eta^{\mu\alpha}\eta^{\nu\sigma}\eta^{\beta\gamma} + k_1\cdot k_2\eta^{\nu\beta}\eta^{\mu\sigma}\eta^{\alpha\gamma} + k_2\cdot k_3\eta^{\nu\beta}\eta^{\mu\sigma}\eta^{\alpha\gamma}\right. \nonumber \\
&& \left.\null+k_2\cdot k_3\eta^{\sigma\gamma}\eta^{\mu\nu}\eta^{\alpha\beta} + k_3\cdot k_1\eta^{\sigma\gamma}\eta^{\mu\nu}\eta^{\alpha\beta} + k_3\cdot k_1\eta^{\mu\alpha}\eta^{\nu\sigma}\eta^{\beta\gamma} \right)\nonumber \\
&& \null+ 2\left(k_1^\nu k_1^\gamma \eta^{\mu\alpha}\eta^{\beta\sigma} + k_2^\sigma k_2^\mu \eta^{\nu\beta}\eta^{\gamma\mu} + k_3^\mu k_3^\beta \eta^{\sigma\gamma}\eta^{\alpha\nu}\right) \nonumber \\
&& \null- \left(k_1^\beta k_2^\mu \eta^{\alpha\nu}\eta^{\sigma\gamma} + k_2^\gamma k_3^\nu \eta^{\beta\sigma}\eta^{\mu\alpha} + k_3^\beta k_1^\mu \eta^{\gamma\mu}\eta^{\nu\beta}\right) \nonumber \\
&& \null+  \left(k_1^\sigma k_2^\gamma\eta^{\mu\nu}\eta^{\alpha\beta} + k_2^\mu k_3^\alpha\eta^{\nu\sigma}\eta^{\beta\gamma} + k_3^\nu k_1^\beta\eta^{\sigma\mu}\eta^{\gamma\alpha}\right) \nonumber \\
&& \null+ \left(k_1^\sigma k_1^\gamma\eta^{\mu\nu}\eta^{\alpha\beta} + k_1^\nu k_1^\beta\eta^{\mu\sigma}\eta^{\alpha\gamma} + k_2^\mu k_2^\alpha\eta^{\nu\sigma}\eta^{\beta\gamma}\right. \nonumber \\
&& \left. \null+  k_2^\sigma k_2^\gamma\eta^{\nu\mu}\eta^{\gamma\alpha} + k_3^\nu k_3^\beta\eta^{\sigma\mu}\eta^{\gamma\alpha} + k_3^\mu k_3^\alpha\eta^{\sigma\nu}\eta^{\alpha\beta}\right)\nonumber \\
&& \null+ 2 \left(k_1^\nu k_2^\gamma\eta^{\beta\mu}\eta^{\alpha\sigma} + k_1^\mu k_2^\gamma\eta^{\alpha\nu}\eta^{\beta\sigma} + k_2^\sigma k_3^\alpha\eta^{\gamma\nu}\eta^{\beta\mu} \right. \nonumber \\
&& \left.\null + k_2^\nu k_3^\alpha\eta^{\beta\sigma}\eta^{\gamma\mu} + k_3^\mu k_1^\beta\eta^{\alpha\sigma}\eta^{\gamma\nu} + k_3^\sigma k_1^\beta\eta^{\gamma\mu}\eta^{\alpha\sigma} \right) \nonumber \\
&& \null+ 2\left(k_1^\nu k_2^\mu\eta^{\beta\sigma}\eta^{\gamma\alpha} + k_2^\sigma k_3^\nu\eta^{\gamma\mu}\eta^{\alpha\beta} + k_3^\mu k_1^\sigma\eta^{\alpha\nu}\eta^{\beta\gamma} \right)  \\
&& \left.\null - 2\left(k_1\cdot k_2 \eta^{\alpha\nu}\eta^{\beta\sigma}\eta^{\gamma\mu} + k_2\cdot k_3 \eta^{\beta\sigma}\eta^{\gamma\mu}\eta^{\alpha\nu} + k_3\cdot k_1 \eta^{\gamma\mu}\eta^{\alpha\nu}\eta^{\gamma\sigma}\right)\frac{}{}\right]. \nonumber
\end{eqnarray}

To find the radiative field contribution from this three-point vertex, two instances of the lowest-order field solution will be stitched together with this vertex to find a higher order contribution. The lowest-order field in momentum space is given by
\begin{equation}
h^{\rho\sigma}(l_\alpha) = \frac{\kappa}{2}\sum_\alpha^N m_\alpha\frac{e^{il_\alpha\cdot b_\alpha}}{l_\alpha^2}(2\pi)\delta(l_\alpha\cdot v_\alpha)\left[v_\alpha^\rho v_\alpha^\sigma - \frac{\eta^{\rho\sigma}}{d-2}\right].
\end{equation}
The three-point vertex allows for a purely gravitational source to be found, which corresponds to a component of the pseudotensor, $t^{\mu\nu}$. This component of the source that generates radiation is given by
\begin{equation}
t^{\sigma\lambda}(k) = \frac{1}{2}\int_{l_\alpha,l_\beta} V^{\mu\rho,\nu\tau,\sigma\lambda}(-l_\alpha,-l_\beta,k)h_{\mu\rho}(l_\alpha)h_{\nu\tau}(l_\beta)\delta^d(k-l_\alpha-l_\beta).
\end{equation}
Since the lowest-order solutions for $l_\alpha$ and $l_\beta$ are symmetric, the symmetrization is only needed for indices $\sigma$ and $\lambda$ in $V^{\mu\rho,\nu\tau,\sigma\lambda}$. Focusing on the integrand and breaking down the two lowest-order solutions into four terms gives
\begin{equation}
h^{\mu\rho}(l_\alpha)h^{\nu\tau}(l_\beta) \propto \left[v_\alpha^\mu v_\alpha^\rho v_\beta^\nu v_\beta^\tau - \frac{1}{d-2}\left(v_\alpha^\mu v_\alpha^\rho \eta^{\nu\tau} + v_\beta^\nu v_\beta^\tau \eta^{\mu\rho} \right) + \frac{1}{(d-2)^2}\left(\eta^{\mu\rho}\eta^{\nu\tau}\right)\right].
\end{equation}
Using Mathematica to perform the index contractions gives
\begin{eqnarray}
(2c)^{\sigma\tau}(k) &=& \left(\frac{\kappa}{2}\right)^2\sum_{\substack{\alpha=1\\ \beta\neq\alpha}}^N m_\alpha m_\beta\int_{l_\alpha,l_\beta}\mu_{\alpha,\beta}(k)\left[  2v_\alpha^\sigma v_\alpha^\lambda\left((k\cdot v_\beta)^2-\frac{l_\alpha^2}{d-2}\right)  \right. \\
&& \null + \left(v_\alpha^\sigma v_\beta^\lambda + v_\alpha^\lambda v_\beta^\sigma\right) \left(l_\alpha^2 v_\alpha\cdot v_\beta - k\cdot v_\alpha k\cdot v_\beta\right) -2 \left(v_\alpha^\sigma l_\alpha^\lambda + v_\alpha^\lambda l_\alpha^\sigma\right)\left(v_\alpha\cdot v_\beta k\cdot v_\beta\right)  \nonumber \\
&& \left.\null + l_\alpha^\sigma l_\alpha^\lambda\left((v_\alpha\cdot v_\beta)^2-\frac{1}{d-2}\right)  +\eta^{\sigma\lambda}\left(k\cdot v_\alpha k\cdot v_\beta v_\alpha\cdot v_\beta -\frac{l_\alpha^2}{2}\left((v_\alpha\cdot v_\beta)^2-\frac{1}{d-2}\right)\right) \right], \nonumber 
\end{eqnarray}
where this result gives the integrand of the diagram (2c), as shown in Eq.~\eqref{2c}.

For calculating the additional gravitational radiation diagrams due to Yang-Mills contributions, the Feynman rules for scattering outlined by Rodigast's thesis give the necessary three-point vertex~\cite{Ebert:2007gf,RodigastThesis}. The Feynman rule for the three-point vertex with two gluons and one graviton can be found from the interaction term in the Lagrangian,
\begin{eqnarray}
\mathcal{L} &=& \sqrt{-g} g^{\mu\rho}g^{\nu\sigma} \partial_\mu A_\nu^a \partial_{[\rho}A_{\sigma]}^a + \dots \nonumber \\
&\approx& \kappa \left(\eta^{\mu\tau}\eta^{\rho\lambda}\eta^{\nu\sigma} + \eta^{\mu\rho}\eta^{\nu\tau}\eta^{\sigma\lambda} - \frac{1}{2} \eta^{\tau\lambda}\eta^{\mu\rho}\eta^{\nu\sigma}\right)h_{\tau\lambda}\partial_\mu A_\nu^a \partial_{[\rho}A_{\sigma]}^a.
\end{eqnarray}
Taking the functional derivatives and properly symmetrizing over all indices and momenta gives
\begin{eqnarray}
\Gamma^{\tau\lambda,\mu \tilde{a},\nu \tilde{b}}(k,p,q) &=& -2i\delta^{\tilde{a}\tilde{b}}\left(p^{(\tau}q^{\lambda)}\eta^{\mu\nu} + \frac{1}{2}p\cdot q (\eta^{\tau\mu}\eta^{\lambda\nu} \eta^{\tau\nu}\eta^{\lambda\mu}-\eta^{\tau\lambda}\eta^{\mu\nu}) \right.\nonumber\\
&& \left.\null+ \frac{1}{2}\eta^{\tau\lambda}p^\nu q^\mu - q^\mu \eta^{\nu(\lambda}p^{\tau)} - p^\nu \eta^{\mu(\tau}q^{\lambda)}\right),
\end{eqnarray}
where a factor of $2/\kappa$ was added to have the same conventions as DeWitt's three-point vertex. This allows us to use the same formula for calculating the contribution to the radiation source. By reusing the lowest-order result for $A^{\mu a}(l)|_{\mathcal{O}(g^1)}$ and switching $a \rightarrow \tilde{a}$, the solution to diagram (2i) is
\begin{equation}
(\mbox{2i})^{\mu\nu}(k) = \frac{1}{2}\int_{l_\alpha,l_\beta}i\Gamma^{\mu\nu,\rho\tilde{a},\sigma\tilde{b}}(- k, l_\alpha,l_\beta)A^{\tilde{a}}_\rho(l_\alpha)|_{\mathcal{O}(g^1)}A^{\tilde{b}}_\sigma(l_\beta)|_{\mathcal{O}(g^1)}\delta^d(k-l_\alpha-l_\beta).
\end{equation}
Plugging in the lowest-order solution gives
\begin{eqnarray}
(\mbox{2i})^{\mu\nu}(k) &=& \tilde{g}^2\sum_{\substack{\alpha=1\\ \beta\neq\alpha}}^N\int_{l_\alpha,l_\beta}\mu_{\alpha,\beta}(k)c_\alpha^{\tilde{a}}c_\beta^{\tilde{a}}\left[ \frac{1}{2}(v_\alpha^\mu v_\beta^\nu + v_\alpha^\nu v_\beta^\mu - \eta^{\mu\nu}v_\alpha\cdot v_\beta)l_\alpha\cdot l_\beta \right. \nonumber \\
&& \left.\null+ v_\alpha\cdot v_\beta l_\alpha^{(\mu}l_\beta^{\nu)}+\frac{1}{2}\eta^{\mu\nu}k\cdot v_\alpha k\cdot v_\beta - k\cdot v_\beta v_\alpha^{(\mu}l_\beta^{\nu)} - k\cdot v_\alpha v_\beta^{(\mu}l_\alpha^{\nu)}\right],
\end{eqnarray}
which can be shown to agree with the algebraic result found in Eq.~\eqref{EYMcontrib}.

\bibliographystyle{utphys}
\bibliography{EYMrad3.bbl}

\providecommand{\href}[2]{#2}\begingroup\raggedright\begin{thebibliography}{10}

\bibitem{Kawai:1985xq}
H.~Kawai, D.~C. Lewellen, and S.~H.~H. Tye, ``{A Relation Between Tree
  Amplitudes of Closed and Open Strings},''
\href{http://dx.doi.org/10.1016/0550-3213(86)90362-7}{{\em Nucl. Phys.}
  {\bfseries B269} (1986) 1--23}.

\bibitem{Bern:2008qj}
Z.~Bern, J.~J.~M. Carrasco, and H.~Johansson, ``{New Relations for Gauge-Theory
  Amplitudes},'' \href{http://dx.doi.org/10.1103/PhysRevD.78.085011}{{\em Phys.
  Rev.} {\bfseries D78} (2008) 085011},
\href{http://arxiv.org/abs/0805.3993}{{\ttfamily arXiv:0805.3993 [hep-ph]}}.

\bibitem{Bern:2010yg}
Z.~Bern, T.~Dennen, Y.-t. Huang, and M.~Kiermaier, ``{Gravity as the Square of
  Gauge Theory},'' \href{http://dx.doi.org/10.1103/PhysRevD.82.065003}{{\em
  Phys. Rev.} {\bfseries D82} (2010) 065003},
\href{http://arxiv.org/abs/1004.0693}{{\ttfamily arXiv:1004.0693 [hep-th]}}.

\bibitem{Bern:2010ue}
Z.~Bern, J.~J.~M. Carrasco, and H.~Johansson, ``{Perturbative Quantum Gravity
  as a Double Copy of Gauge Theory},''
  \href{http://dx.doi.org/10.1103/PhysRevLett.105.061602}{{\em Phys. Rev.
  Lett.} {\bfseries 105} (2010) 061602},
\href{http://arxiv.org/abs/1004.0476}{{\ttfamily arXiv:1004.0476 [hep-th]}}.

\bibitem{ArkaniHamed:2009dn}
N.~Arkani-Hamed, F.~Cachazo, C.~Cheung, and J.~Kaplan, ``{A Duality For The S
  Matrix},'' \href{http://dx.doi.org/10.1007/JHEP03(2010)020}{{\em JHEP}
  {\bfseries 03} (2010) 020},
\href{http://arxiv.org/abs/0907.5418}{{\ttfamily arXiv:0907.5418 [hep-th]}}.

\bibitem{Chen:2011jxa}
Y.-X. Chen, Y.-J. Du, and B.~Feng, ``{A Proof of the Explicit Minimal-basis
  Expansion of Tree Amplitudes in Gauge Field Theory},''
  \href{http://dx.doi.org/10.1007/JHEP02(2011)112}{{\em JHEP} {\bfseries 02}
  (2011) 112},
\href{http://arxiv.org/abs/1101.0009}{{\ttfamily arXiv:1101.0009 [hep-th]}}.

\bibitem{delaCruz:2015dpa}
L.~de~la Cruz, A.~Kniss, and S.~Weinzierl, ``{Proof of the fundamental BCJ
  relations for QCD amplitudes},''
  \href{http://dx.doi.org/10.1007/JHEP09(2015)197}{{\em JHEP} {\bfseries 09}
  (2015) 197},
\href{http://arxiv.org/abs/1508.01432}{{\ttfamily arXiv:1508.01432 [hep-th]}}.

\bibitem{Carrasco:2011hw}
J.~J.~M. Carrasco and H.~Johansson, ``{Generic multiloop methods and
  application to N=4 super-Yang-Mills},''
  \href{http://dx.doi.org/10.1088/1751-8113/44/45/454004}{{\em J. Phys.}
  {\bfseries A44} (2011) 454004},
\href{http://arxiv.org/abs/1103.3298}{{\ttfamily arXiv:1103.3298 [hep-th]}}.

\bibitem{Carrasco:2011mn}
J.~J. Carrasco and H.~Johansson, ``{Five-Point Amplitudes in N=4
  Super-Yang-Mills Theory and N=8 Supergravity},''
  \href{http://dx.doi.org/10.1103/PhysRevD.85.025006}{{\em Phys. Rev.}
  {\bfseries D85} (2012) 025006},
\href{http://arxiv.org/abs/1106.4711}{{\ttfamily arXiv:1106.4711 [hep-th]}}.

\bibitem{Bern:2012uf}
Z.~Bern, J.~J.~M. Carrasco, L.~J. Dixon, H.~Johansson, and R.~Roiban,
  ``{Simplifying Multiloop Integrands and Ultraviolet Divergences of Gauge
  Theory and Gravity Amplitudes},''
  \href{http://dx.doi.org/10.1103/PhysRevD.85.105014}{{\em Phys. Rev.}
  {\bfseries D85} (2012) 105014},
\href{http://arxiv.org/abs/1201.5366}{{\ttfamily arXiv:1201.5366 [hep-th]}}.

\bibitem{Bern:2013yya}
Z.~Bern, S.~Davies, T.~Dennen, Y.-t. Huang, and J.~Nohle, ``{Color-Kinematics
  Duality for Pure Yang-Mills and Gravity at One and Two Loops},''
  \href{http://dx.doi.org/10.1103/PhysRevD.92.045041}{{\em Phys. Rev.}
  {\bfseries D92} no.~4, (2015) 045041},
\href{http://arxiv.org/abs/1303.6605}{{\ttfamily arXiv:1303.6605 [hep-th]}}.

\bibitem{BjerrumBohr:2009rd}
N.~E.~J. Bjerrum-Bohr, P.~H. Damgaard, and P.~Vanhove, ``{Minimal Basis for
  Gauge Theory Amplitudes},''
  \href{http://dx.doi.org/10.1103/PhysRevLett.103.161602}{{\em Phys. Rev.
  Lett.} {\bfseries 103} (2009) 161602},
\href{http://arxiv.org/abs/0907.1425}{{\ttfamily arXiv:0907.1425 [hep-th]}}.

\bibitem{Stieberger:2009hq}
S.~Stieberger, ``{Open \& Closed vs. Pure Open String Disk Amplitudes},''
\href{http://arxiv.org/abs/0907.2211}{{\ttfamily arXiv:0907.2211 [hep-th]}}.

\bibitem{BjerrumBohr:2010zs}
N.~E.~J. Bjerrum-Bohr, P.~H. Damgaard, T.~Sondergaard, and P.~Vanhove,
  ``{Monodromy and Jacobi-like Relations for Color-Ordered Amplitudes},''
  \href{http://dx.doi.org/10.1007/JHEP06(2010)003}{{\em JHEP} {\bfseries 06}
  (2010) 003},
\href{http://arxiv.org/abs/1003.2403}{{\ttfamily arXiv:1003.2403 [hep-th]}}.

\bibitem{Tye:2010dd}
S.~H. Henry~Tye and Y.~Zhang, ``{Dual Identities inside the Gluon and the
  Graviton Scattering Amplitudes},''
  \href{http://dx.doi.org/10.1007/JHEP06(2010)071,
  10.1007/JHEP04(2011)114}{{\em JHEP} {\bfseries 06} (2010) 071},
  \href{http://arxiv.org/abs/1003.1732}{{\ttfamily arXiv:1003.1732 [hep-th]}}.
[Erratum: JHEP04,114(2011)].

\bibitem{He:2016mzd}
S.~He and O.~Schlotterer, ``{New Relations for Gauge-Theory and Gravity
  Amplitudes at Loop Level},''
  \href{http://dx.doi.org/10.1103/PhysRevLett.118.161601}{{\em Phys. Rev.
  Lett.} {\bfseries 118} no.~16, (2017) 161601},
\href{http://arxiv.org/abs/1612.00417}{{\ttfamily arXiv:1612.00417 [hep-th]}}.

\bibitem{He:2017spx}
S.~He, O.~Schlotterer, and Y.~Zhang, ``{New BCJ representations for one-loop
  amplitudes in gauge theories and gravity},''
\href{http://arxiv.org/abs/1706.00640}{{\ttfamily arXiv:1706.00640 [hep-th]}}.

\bibitem{Adamo:2015gia}
T.~Adamo, E.~Casali, K.~A. Roehrig, and D.~Skinner, ``{On tree amplitudes of
  supersymmetric Einstein-Yang-Mills theory},''
  \href{http://dx.doi.org/10.1007/JHEP12(2015)177}{{\em JHEP} {\bfseries 12}
  (2015) 177},
\href{http://arxiv.org/abs/1507.02207}{{\ttfamily arXiv:1507.02207 [hep-th]}}.

\bibitem{delaCruz:2016gnm}
L.~de~la Cruz, A.~Kniss, and S.~Weinzierl, ``{Relations for
  Einstein–Yang–Mills amplitudes from the CHY representation},''
  \href{http://dx.doi.org/10.1016/j.physletb.2017.01.036}{{\em Phys. Lett.}
  {\bfseries B767} (2017) 86--90},
\href{http://arxiv.org/abs/1607.06036}{{\ttfamily arXiv:1607.06036 [hep-th]}}.

\bibitem{Stieberger:2016lng}
S.~Stieberger and T.~R. Taylor, ``{New relations for Einstein–Yang–Mills
  amplitudes},'' \href{http://dx.doi.org/10.1016/j.nuclphysb.2016.09.014}{{\em
  Nucl. Phys.} {\bfseries B913} (2016) 151--162},
\href{http://arxiv.org/abs/1606.09616}{{\ttfamily arXiv:1606.09616 [hep-th]}}.

\bibitem{Du:2016wkt}
Y.-J. Du, F.~Teng, and Y.-S. Wu, ``{Direct Evaluation of $n$-point single-trace
  MHV amplitudes in 4d Einstein-Yang-Mills theory using the CHY Formalism},''
  \href{http://dx.doi.org/10.1007/JHEP09(2016)171}{{\em JHEP} {\bfseries 09}
  (2016) 171},
\href{http://arxiv.org/abs/1608.00883}{{\ttfamily arXiv:1608.00883 [hep-th]}}.

\bibitem{Cachazo:2014nsa}
F.~Cachazo, S.~He, and E.~Y. Yuan, ``{Einstein-Yang-Mills Scattering Amplitudes
  From Scattering Equations},''
  \href{http://dx.doi.org/10.1007/JHEP01(2015)121}{{\em JHEP} {\bfseries 01}
  (2015) 121},
\href{http://arxiv.org/abs/1409.8256}{{\ttfamily arXiv:1409.8256 [hep-th]}}.

\bibitem{Nandan:2016pya}
D.~Nandan, J.~Plefka, O.~Schlotterer, and C.~Wen, ``{Einstein-Yang-Mills from
  pure Yang-Mills amplitudes},''
  \href{http://dx.doi.org/10.1007/JHEP10(2016)070}{{\em JHEP} {\bfseries 10}
  (2016) 070},
\href{http://arxiv.org/abs/1607.05701}{{\ttfamily arXiv:1607.05701 [hep-th]}}.

\bibitem{Schlotterer:2016cxa}
O.~Schlotterer, ``{Amplitude relations in heterotic string theory and
  Einstein-Yang-Mills},'' \href{http://dx.doi.org/10.1007/JHEP11(2016)074}{{\em
  JHEP} {\bfseries 11} (2016) 074},
\href{http://arxiv.org/abs/1608.00130}{{\ttfamily arXiv:1608.00130 [hep-th]}}.

\bibitem{Cachazo:2013gna}
F.~Cachazo, S.~He, and E.~Y. Yuan, ``{Scattering equations and
  Kawai-Lewellen-Tye orthogonality},''
  \href{http://dx.doi.org/10.1103/PhysRevD.90.065001}{{\em Phys. Rev.}
  {\bfseries D90} no.~6, (2014) 065001},
\href{http://arxiv.org/abs/1306.6575}{{\ttfamily arXiv:1306.6575 [hep-th]}}.

\bibitem{White:2016jzc}
C.~D. White, ``{Exact solutions for the biadjoint scalar field},''
\href{http://arxiv.org/abs/1606.04724}{{\ttfamily arXiv:1606.04724 [hep-th]}}.

\bibitem{Chiodaroli:2014xia}
M.~Chiodaroli, M.~Günaydin, H.~Johansson, and R.~Roiban, ``{Scattering
  amplitudes in $ \mathcal{N}=2 $ Maxwell-Einstein and Yang-Mills/Einstein
  supergravity},'' \href{http://dx.doi.org/10.1007/JHEP01(2015)081}{{\em JHEP}
  {\bfseries 01} (2015) 081},
\href{http://arxiv.org/abs/1408.0764}{{\ttfamily arXiv:1408.0764 [hep-th]}}.

\bibitem{Chiodaroli:2015rdg}
M.~Chiodaroli, M.~Gunaydin, H.~Johansson, and R.~Roiban, ``{Spontaneously
  Broken Yang-Mills-Einstein Supergravities as Double Copies},''
  \href{http://dx.doi.org/10.1007/JHEP06(2017)064}{{\em JHEP} {\bfseries 06}
  (2017) 064},
\href{http://arxiv.org/abs/1511.01740}{{\ttfamily arXiv:1511.01740 [hep-th]}}.

\bibitem{Chiodaroli:2016jqw}
M.~Chiodaroli, ``{Simplifying amplitudes in Maxwell-Einstein and
  Yang-Mills-Einstein supergravities},''
\newblock 2016.
\newblock \href{http://arxiv.org/abs/1607.04129}{{\ttfamily arXiv:1607.04129
  [hep-th]}}.
\newblock
\url{https://inspirehep.net/record/1475711/files/arXiv:1607.04129.pdf}.
\newblock

\bibitem{PhysRevLett.116.061102}
{\bfseries LIGO Scientific Collaboration and Virgo Collaboration}
  Collaboration, B.~P. Abbott, R.~Abbott, T.~D. Abbott, {\em et~al.},
  ``Observation of gravitational waves from a binary black hole merger,''
  \href{http://dx.doi.org/10.1103/PhysRevLett.116.061102}{{\em Phys. Rev.
  Lett.} {\bfseries 116} (Feb, 2016) 061102}.
  \url{http://link.aps.org/doi/10.1103/PhysRevLett.116.061102}.

\bibitem{Anastasiou:2013hba}
A.~Anastasiou, L.~Borsten, M.~J. Duff, L.~J. Hughes, and S.~Nagy, ``{A magic
  pyramid of supergravities},''
  \href{http://dx.doi.org/10.1007/JHEP04(2014)178}{{\em JHEP} {\bfseries 04}
  (2014) 178},
\href{http://arxiv.org/abs/1312.6523}{{\ttfamily arXiv:1312.6523 [hep-th]}}.

\bibitem{Anastasiou:2017nsz}
A.~Anastasiou, L.~Borsten, M.~J. Duff, {\em et~al.}, ``{Are all supergravity
  theories Yang-Mills squared?},''
\href{http://arxiv.org/abs/1707.03234}{{\ttfamily arXiv:1707.03234 [hep-th]}}.

\bibitem{Johansson:2017srf}
H.~Johansson and J.~Nohle, ``{Conformal Gravity from Gauge Theory},''
\href{http://arxiv.org/abs/1707.02965}{{\ttfamily arXiv:1707.02965 [hep-th]}}.

\bibitem{Monteiro:2014cda}
R.~Monteiro, D.~O'Connell, and C.~D. White, ``{Black holes and the double
  copy},'' \href{http://dx.doi.org/10.1007/JHEP12(2014)056}{{\em JHEP}
  {\bfseries 12} (2014) 056},
\href{http://arxiv.org/abs/1410.0239}{{\ttfamily arXiv:1410.0239 [hep-th]}}.

\bibitem{Ridgway:2015fdl}
A.~K. Ridgway and M.~B. Wise, ``{Static Spherically Symmetric Kerr-Schild
  Metrics and Implications for the Classical Double Copy},''
\href{http://arxiv.org/abs/1512.02243}{{\ttfamily arXiv:1512.02243 [hep-th]}}.

\bibitem{Luna:2015paa}
A.~Luna, R.~Monteiro, D.~O'Connell, and C.~D. White, ``{The classical double
  copy for Taub–NUT spacetime},''
  \href{http://dx.doi.org/10.1016/j.physletb.2015.09.021}{{\em Phys. Lett.}
  {\bfseries B750} (2015) 272--277},
\href{http://arxiv.org/abs/1507.01869}{{\ttfamily arXiv:1507.01869 [hep-th]}}.

\bibitem{Luna:2016due}
A.~Luna, R.~Monteiro, I.~Nicholson, D.~O'Connell, and C.~D. White, ``{The
  double copy: Bremsstrahlung and accelerating black holes},''
  \href{http://dx.doi.org/10.1007/JHEP06(2016)023}{{\em JHEP} {\bfseries 06}
  (2016) 023},
\href{http://arxiv.org/abs/1603.05737}{{\ttfamily arXiv:1603.05737 [hep-th]}}.

\bibitem{Bahjat-Abbas:2017htu}
N.~Bahjat-Abbas, A.~Luna, and C.~D. White, ``{The Kerr-Schild double copy in
  curved spacetime},''
\href{http://arxiv.org/abs/1710.01953}{{\ttfamily arXiv:1710.01953 [hep-th]}}.

\bibitem{Adamo:2017nia}
T.~Adamo, E.~Casali, L.~Mason, and S.~Nekovar, ``{Scattering on plane waves and
  the double copy},''
\href{http://arxiv.org/abs/1706.08925}{{\ttfamily arXiv:1706.08925 [hep-th]}}.

\bibitem{Luna:2017dtq}
A.~Luna, I.~Nicholson, D.~O'Connell, and C.~D. White, ``{Inelastic Black Hole
  Scattering from Charged Scalar Amplitudes},''
\href{http://arxiv.org/abs/1711.03901}{{\ttfamily arXiv:1711.03901 [hep-th]}}.

\bibitem{Goldberger:2016iau}
W.~D. Goldberger and A.~K. Ridgway, ``{Radiation and the classical double copy
  for color charges},''
  \href{http://dx.doi.org/10.1103/PhysRevD.95.125010}{{\em Phys. Rev.}
  {\bfseries D95} no.~12, (2017) 125010},
\href{http://arxiv.org/abs/1611.03493}{{\ttfamily arXiv:1611.03493 [hep-th]}}.

\bibitem{Goldberger:2017frp}
W.~D. Goldberger, S.~G. Prabhu, and J.~O. Thompson, ``{Classical gluon and
  graviton radiation from the bi-adjoint scalar double copy},''
  \href{http://dx.doi.org/10.1103/PhysRevD.96.065009}{{\em Phys. Rev.}
  {\bfseries D96} no.~6, (2017) 065009},
\href{http://arxiv.org/abs/1705.09263}{{\ttfamily arXiv:1705.09263 [hep-th]}}.

\bibitem{MTW}
C.~Misner, K.~Thorne, and J.~Wheeler, {\em Gravitation}.
\newblock No.~pt. 3 in Gravitation. W. H. Freeman, 1973.
\newblock \url{https://books.google.com/books?id=w4Gigq3tY1kC}.

\bibitem{Chu:2016ngc}
Y.-Z. Chu, ``{More On Cosmological Gravitational Waves And Their Memories},''
  \href{http://dx.doi.org/10.1088/1361-6382/aa8392}{{\em Class. Quant. Grav.}
  {\bfseries 34} no.~19, (2017) 194001},
\href{http://arxiv.org/abs/1611.00018}{{\ttfamily arXiv:1611.00018 [gr-qc]}}.

\bibitem{Will:1996zj}
C.~M. Will and A.~G. Wiseman, ``{Gravitational radiation from compact binary
  systems: Gravitational wave forms and energy loss to second postNewtonian
  order},'' \href{http://dx.doi.org/10.1103/PhysRevD.54.4813}{{\em Phys. Rev.}
  {\bfseries D54} (1996) 4813--4848},
\href{http://arxiv.org/abs/gr-qc/9608012}{{\ttfamily arXiv:gr-qc/9608012
  [gr-qc]}}.

\bibitem{Flanagan:2005yc}
E.~E. Flanagan and S.~A. Hughes, ``{The Basics of gravitational wave theory},''
  \href{http://dx.doi.org/10.1088/1367-2630/7/1/204}{{\em New J. Phys.}
  {\bfseries 7} (2005) 204},
\href{http://arxiv.org/abs/gr-qc/0501041}{{\ttfamily arXiv:gr-qc/0501041
  [gr-qc]}}.

\bibitem{Blanchet}
L.~{Blanchet}, ``{Gravitational Radiation from Post-Newtonian Sources and
  Inspiralling Compact Binaries},''
  \href{http://dx.doi.org/10.12942/lrr-2014-2}{{\em Living Reviews in
  Relativity} {\bfseries 17} (Feb., 2014) 2},
  \href{http://arxiv.org/abs/1310.1528}{{\ttfamily arXiv:1310.1528 [gr-qc]}}.

\bibitem{Porto:2017dgs}
R.~A. Porto and I.~Z. Rothstein, ``{Apparent ambiguities in the post-Newtonian
  expansion for binary systems},''
  \href{http://dx.doi.org/10.1103/PhysRevD.96.024062}{{\em Phys. Rev.}
  {\bfseries D96} no.~2, (2017) 024062},
\href{http://arxiv.org/abs/1703.06433}{{\ttfamily arXiv:1703.06433 [gr-qc]}}.

\bibitem{Bernard:2017ktp}
L.~Bernard, L.~Blanchet, G.~Faye, and T.~Marchand, ``{Center-of-Mass Equations
  of Motion and Conserved Integrals of Compact Binary Systems at the Fourth
  Post-Newtonian Order},''
\href{http://arxiv.org/abs/1711.00283}{{\ttfamily arXiv:1711.00283 [gr-qc]}}.

\bibitem{Bohe:2016gbl}
A.~Bohé {\em et~al.}, ``{Improved effective-one-body model of spinning,
  nonprecessing binary black holes for the era of gravitational-wave
  astrophysics with advanced detectors},''
  \href{http://dx.doi.org/10.1103/PhysRevD.95.044028}{{\em Phys. Rev.}
  {\bfseries D95} no.~4, (2017) 044028},
\href{http://arxiv.org/abs/1611.03703}{{\ttfamily arXiv:1611.03703 [gr-qc]}}.

\bibitem{Moxon:2017ozd}
J.~Moxon and a.~Flanagan, ``{Radiation-Reaction Force on a Small Charged Body
  to Second Order},''
\href{http://arxiv.org/abs/1711.05212}{{\ttfamily arXiv:1711.05212 [gr-qc]}}.

\bibitem{Damour:2016gwp}
T.~Damour, ``{Gravitational scattering, post-Minkowskian approximation and
  Effective One-Body theory},''
  \href{http://dx.doi.org/10.1103/PhysRevD.94.104015}{{\em Phys. Rev.}
  {\bfseries D94} no.~10, (2016) 104015},
\href{http://arxiv.org/abs/1609.00354}{{\ttfamily arXiv:1609.00354 [gr-qc]}}.

\bibitem{Bini:2017xzy}
D.~Bini and T.~Damour, ``{Gravitational spin-orbit coupling in binary systems,
  post-Minkowskian approximation and effective one-body theory},''
  \href{http://dx.doi.org/10.1103/PhysRevD.96.104038}{{\em Phys. Rev.}
  {\bfseries D96} no.~10, (2017) 104038},
\href{http://arxiv.org/abs/1709.00590}{{\ttfamily arXiv:1709.00590 [gr-qc]}}.

\bibitem{Damour:2017zjx}
T.~Damour, ``{High-energy gravitational scattering and the general relativistic
  two-body problem},''
\href{http://arxiv.org/abs/1710.10599}{{\ttfamily arXiv:1710.10599 [gr-qc]}}.

\bibitem{Sikivie}
P.~Sikivie and N.~Weiss, ``Classical yang-mills theory in the presence of
  external sources,'' \href{http://dx.doi.org/10.1103/PhysRevD.18.3809}{{\em
  Phys. Rev. D} {\bfseries 18} (Nov, 1978) 3809--3821}.
  \url{https://link.aps.org/doi/10.1103/PhysRevD.18.3809}.

\bibitem{jackson}
J.~D. Jackson, {\em Classical electrodynamics}.
\newblock Wiley, New York, {NY}, 3rd ed.~ed., 1999.
\newblock \url{http://cdsweb.cern.ch/record/490457}.

\bibitem{Dirac}
P.~Dirac, {\em General Theory of Relativity}.
\newblock Physics Notes. Princeton University Press, 1975.
\newblock \url{https://books.google.com/books?id=qkWPDAAAQBAJ}.

\bibitem{LLpseudo}
L.~Landau and E.~Lifshitz,
  \href{http://dx.doi.org/https://doi.org/10.1016/B978-0-08-025072-4.50018-6}{``Chaper
  11 - the gravitational field equations,''} in {\em The Classical Theory of
  Fields (Fourth Edition)}, L.~Landau and E.~Lifshitz, eds., vol.~2 of {\em
  Course of Theoretical Physics}, pp.~259 -- 294.
\newblock Pergamon, Amsterdam, fourth edition~ed., 1975.
\newblock
  \url{https://www.sciencedirect.com/science/article/pii/B9780080250724500186}.

\bibitem{DeWitt}
B.~S. DeWitt, ``Quantum theory of gravity. iii. applications of the covariant
  theory,'' \href{http://dx.doi.org/10.1103/PhysRev.162.1239}{{\em Phys. Rev.}
  {\bfseries 162} (Oct, 1967) 1239--1256}.
  \url{https://link.aps.org/doi/10.1103/PhysRev.162.1239}.

\bibitem{Sannan}
S.~Sannan, ``Gravity as the limit of the type-ii superstring theory,''
  \href{http://dx.doi.org/10.1103/PhysRevD.34.1749}{{\em Phys. Rev. D}
  {\bfseries 34} (Sep, 1986) 1749--1758}.
  \url{https://link.aps.org/doi/10.1103/PhysRevD.34.1749}.

\bibitem{Ebert:2007gf}
D.~Ebert, J.~Plefka, and A.~Rodigast, ``{Absence of gravitational contributions
  to the running Yang-Mills coupling},''
  \href{http://dx.doi.org/10.1016/j.physletb.2008.01.037}{{\em Phys. Lett.}
  {\bfseries B660} (2008) 579--582},
\href{http://arxiv.org/abs/0710.1002}{{\ttfamily arXiv:0710.1002 [hep-th]}}.

\bibitem{RodigastThesis}
A.~Rodigast, {\em One-Loop Divergences of the Yang-Mills Theory Coupled to
  Gravitation}.
\newblock PhD thesis, Humboldt University, August, 2008.
\newblock
  \url{http://people.physik.hu-berlin.de/~sylvia/qftpha/qft/downloads/DA-andreas.pdf}.

\end{thebibliography}\endgroup

\end{document}